\documentclass[pre,aps,showpacs,notitlepage,twocolumn,superscriptaddress]{revtex4-1}

\usepackage{tabularx}
\usepackage{blindtext}
\usepackage{svg}
\usepackage{float}
\usepackage{mwe}
\usepackage{multirow}
\usepackage{hyperref}
\hypersetup{colorlinks,linkcolor={blue},citecolor={blue},urlcolor={blue}}

\usepackage{graphicx}

\usepackage{bm,amsmath}

\usepackage{latexsym}

\usepackage{verbatim}

\usepackage{xcolor}
\usepackage{subfigure}

\usepackage{gensymb}

\usepackage{soul}

\usepackage{tikz}

\usepackage[T1]{fontenc}
\usepackage{upgreek}
\usepackage{comment}
 \usepackage{fourier}

\newcommand{\romy}[1]{\textcolor{blue}{[Romane: \textbf{#1}]}}
\newcommand{\sara}[1]{\textcolor{red}{[sara: \textbf{#1}]}}
\pdfoutput=1

\begin{document}

\title{Pore network models for the evaporation of complex fluids in porous media}

\author{Romane Le Dizès Castell}
\affiliation{Institute of Physics, University of Amsterdam, Science Park 904, 1098XH Amsterdam, The Netherlands}
\author{Marc Prat}
\affiliation{Institut de Mécanique des Fluides de Toulouse) Allée du Professeur Camille Soula
F-31400 Toulouse, France}
\author{Noushine Shahidzadeh}
\affiliation{Institute of Physics, University of Amsterdam, Science Park 904, 1098XH Amsterdam, The Netherlands}
\author{Sara Jabbari-Farouji}
\altaffiliation[Correspondence to: ]{\hyperlink{s.jabbarifarouji@uva.nl}{s.jabbarifarouji@uva.nl}}
\affiliation{Institute of Physics, University of Amsterdam, Science Park 904, 1098XH Amsterdam, The Netherlands}


\begin{abstract}
Drying of fluids undergoing sol-gel transition in porous media, a process crucial for the consolidation of damaged porous structures in cultural heritage,
often leads to skin formation at the surface. This phenomenon  significantly hinders evaporation, yet its precise impact  on drying kinetics remains poorly understood. To uncover the governing mechanisms, we develop a novel pore network model that closely replicates quasi-2D  experimental porous media in reference~\cite{le_dizes_castell_visualization_2024}, incorporating spatial gradients in pore size distribution, with smaller pores near the evaporation side. We demonstrate that this pore distribution dictates the air invasion path and extends the constant-rate  period of drying in Newtonian liquids, reproducing the experimental  drying curves for water. We further extend our model  to capture the interplay of capillary-driven liquid flows, space-dependent viscosity increases, and localized skin formation. To incorporate skin formation, we implement a viscosity-dependent evaporation pressure rule derived from experimental data on evaporation-induced sol-gel transition within a capillary tube.  We identify a simple relationship: vapor pressure decreases once the meniscus fluid viscosity exceeds a critical threshold. By accounting for localized skin formation through reduced evaporation pressure at high-viscosity throats, our model successfully captures the slowdown of drying kinetics, achieving remarkable agreement with experimental drying curve.

\end{abstract}

\maketitle
\section{Introduction}

Drying of multicomponent fluids is essential to a wide range of industrial or natural processes. It includes the drying of ionic solutions, colloidal suspensions, or polymer solutions undergoing a sol-gel transition, and it can lead to a variety of different physical phenomena such as salt crystallization or skin formation \citep{okuzono_simple_2006,talini_formation_2023, salmon_humidity-insensitive_2017}. A skin can be defined as thin layer near the surface of fluid, with thickness of about 1 to 10 $\mu m$ which has different properties than the bulk material \citep{faiyas_transport_2017}. In the context of colloidal suspensions and polymer solutions skin formation arises from the accumulation of solute particles at the free surface during the drying \citep{brinker_review_1992}. It has been proven experimentally \cite{pauchard_stable_2003, pauchard_buckling_2003, huisman_evaporation_2023}, numerically \cite{cairncross_predicting_1996, okuzono_simple_2006} and analytical approaches \cite{salmon_humidity-insensitive_2017, talini_formation_2023} have demonstrated that solute accumulation at the free surface and the resultant concentration gradient significantly impacts the subsequent drying process, leading to a marked decrease in the evaporation rate.

  A phenomenon similar to skin formation  has been observed in several studies on the drying of multicomponent fluids in capillary porous media, where it  significantly slows  fluid evaporation within pores. For example, in civil engineering, addition of polymeric compounds to mortars has been shown to delay  solidification by reducing evaporation rates~\cite{buhler_dynamics_2013, faiyas_how_2015, faiyas_transport_2017}. Similarly, in soils, the  biofilm formation at the air-water interface through the secretion of extracellular polymeric substances by microorganisms and bacteria \cite{or_extracellular_2007} improves water retention in soils due to strongly reduced evaporation rate~\cite{deng_synergistic_2015, guo_bacterial_2018}. In the field of cultural heritage, recent experimental studies  have demonstrated that fluids undergoing sol-gel transition, used as stone consolidants \cite{wheeler_alkoxysilanes_2005}, can form a skin at the surface of porous media during drying \citep{le_dizes_castell_visualization_2024,le_dizes_castell_sol-gel_2024}. 
\begin{figure}
    \centering
    \includegraphics[width=1\columnwidth]{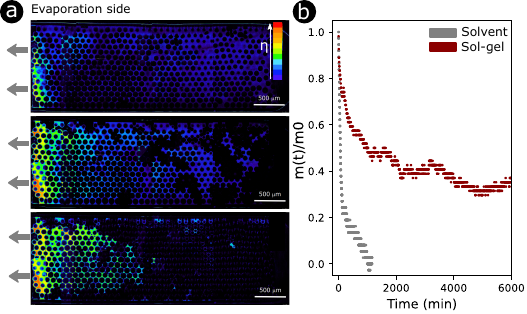}
    \caption{\textbf{Sol-gel transition in a quasi-2D model porous medium}, reproduced from Ref. \cite{le_dizes_castell_visualization_2024}. a) Fluorescence lifetime imaging microscopy confocal images of the polymer solution in porous medium at different stages of the drying during which the viscosity of the fluid varies between 0.05 Pa.s (dark purple) and 2 Pa.s (red). b)The ratio of mass to its initial value $m_0$  over time during the drying process in the porous medium, both for the pure solvent and for the polymer solution undergoing sol-gel transition.}
    \label{fig:experimentalResults}
\end{figure}

To gain direct insights into the skin formation in porous media, \citet{le_dizes_castell_visualization_2024} have studied the sol-gel transition during drying in quasi-2D porous media using fluorescent viscosity-sensitive molecular probes allowing for direct visualization of local viscosity increase within the porous network. The authorsfoundhthat during the sol-gel transition, the viscosity of the fluid increases strongly and inhomogeneously over time, with a stronger increase at the evaporation side, see Fig. \ref{fig:experimentalResults}(a). At the late stages of the drying, the gel is mainly located at the evaporation side of the porous medium. As shown in Fig. \ref{fig:experimentalResults}(b), the drying process is significantly slowed by the sol-gel transition compared to the pure solvent, with the drying rate sharply decreasing from the start of the drying. Ref. \cite{le_dizes_castell_visualization_2024} also demonstrated that the gel forms mainly in small pores. Although this experimental work provides new insights into gel formation within porous media during drying, the  influence of  skin formation  on  drying kinetics  remain to be explained.

Despite wide interest in drying of complex fluids in porous media and their industrial and technological applications, to our knowledge, no theoretical or numerical framework currently explains the evaporation-induced gelation and/or skin formation in porous media. These phenomena result from the interplay of  capillary-driven liquid flows,  viscosity increase during gel formation, and evaporation. 
To delineate the contributions of various factors to the drying process at different stages,  we develop a pore-network model that incorporates the spatio-temporal viscosity changes during the evaporation, using the experimental data of Ref. \cite{le_dizes_castell_visualization_2024} as an input.
The pore-network modeling approach  provides a structured representation of porous media and is based on the invasion percolation algorithm \cite{wilkinson_invasion_1983, wilkinson_percolation_1986}  where the liquid phase in the pore space is invaded by gas phase due to evaporation. Pore network models are used to predict the effect of porous network morphology and the contribution of
different displacement mechanisms. Mass transfer mechanisms at the pore scale include  viscous flow in liquid phase,  capillary pumping
in liquid phase and diffusion in gas phase.

This method has found to be a powerful tool to study the pore-scale drying, as demonstrated by \citet{nowicki_microscopic_1992} and \citet{prat_percolation_1993}. Modern models nowadays incorporate capillarity, gravitational  and viscous effects~\cite{yiotis_2-d_2001, metzger_isothermal_2007}.
In this work, we develop a 2D pore-network model tailored to replicate the quasi-2D microfluidic porous media presented in Fig. \ref{fig:experimentalResults}. Viscosity values during drying are directly extracted from experimental data. The model captures spatio-temporal viscosity variations and implicitly accounts for skin formation by imposing a reduced evaporation pressure when local viscosity exceeds a critical threshold. As such our model is able to predict the drying pattern and kinetics and offers new insights into   the impact of localized skin formation at menisci on overall evaporation rates.

The remainder of this article is organized as follows. In section \ref{sec:Method}, we explain the algorithm of pore network model for isothermal drying and we build the pore network. In section \ref{sec:WaterPNM}, we investigate the drying of a pure liquid in a porous network with spatial gradients in size of throat bodies which enables us to reproduce our experimental results for water evaporation~\cite{le_dizes_castell_visualization_2024}. The influence of a high but uniform viscosity is afterwards investigated in section \ref{sec:PNM_Visco}. In section \ref{sec:SolGelPNM}, we investigate the effects of spatio-temporal gradient of viscosity on the liquid flows during the drying and  accounting for the impact of the skin formation via viscosity-dependent vapor pressure on the drying rate of the porous medium. Finally, we conclude in section~\ref{sec:conclusion} with a summary of key insights and perspectives for future work.

\section{Methods} \label{sec:Method}

In this section, we provide a concise explanation of the building elements of the  isothermal pore network model developed in this work based on the approaches of   Refs \cite{prat_percolation_1993} and \cite{metzger_isothermal_2007}. In our study, gravitational effects are excluded as the  porous network is quasi-2D and evaporation in experimental setting happens horizontally. The capillary effects are taken into account using the invasion percolation algorithm \cite{wilkinson_invasion_1983, wilkinson_percolation_1986} and the transport of vapor by diffusion in the gas phase is implemented. Isothermal drying conditions are considered. As the sol-gel transition is characterized by a strong viscosity increase, we also include the viscous effects following the method of Ref \cite{metzger_isothermal_2007}. 

\subsection{Description of pore network geometry}
\begin{figure}[h!]
    \centering
    \includegraphics[width=0.6\columnwidth]{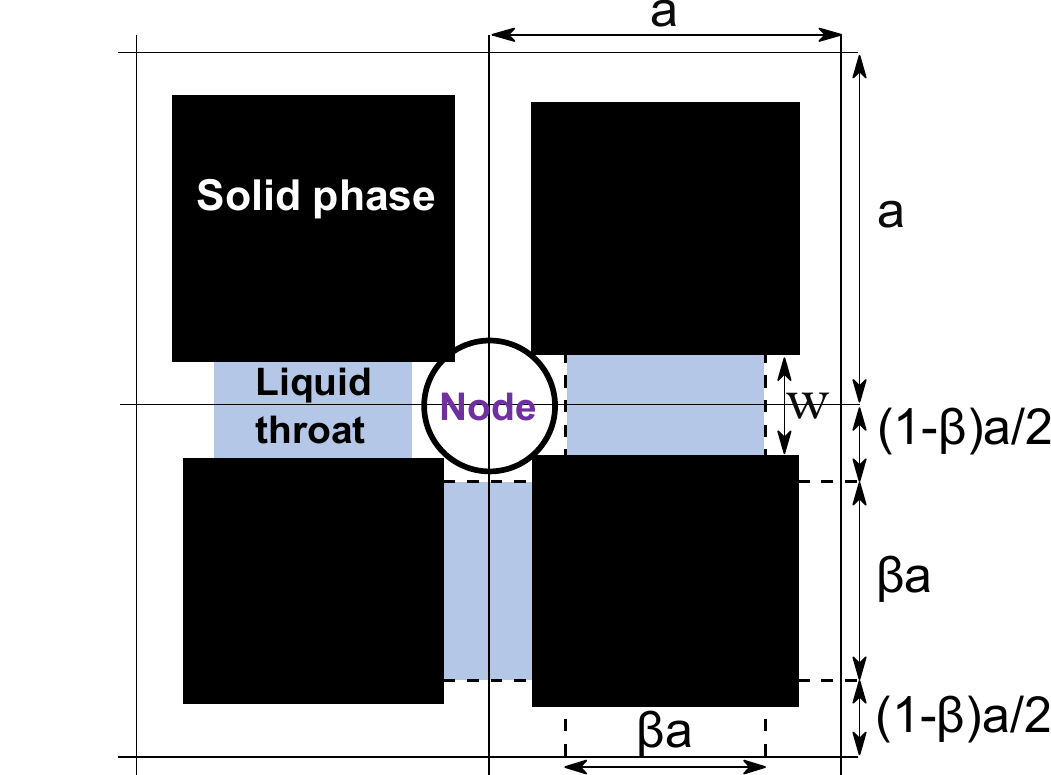}
    \caption{Representation of  basic geometry of  a PN on square lattice, in which three  throats (blue) are filled by liquid, and one  throat is   empty (gaseous) and connected to a gas node in the PN.}
    \label{fig:square}
\end{figure}

The pore network (PN) is represented by a two-dimensional network of nodes bodies connected by  throat bodies with random sizes, see Fig \ref{fig:square}. The node and the throat bodies, hereafter referred to as "nodes" and throats" are positioned at the sites and bonds of a square lattice with side length $a$, respectively. The network has a thickness $e$. 
The space is partitioned into nodes and throats  using the parameter $\beta$ such that the throats have a length of $\beta a$. $\beta$ is a constant coefficient in the interval (0, 1),  which  determines the porosity of the network, see  \cite{prat_isothermal_1995} for details. The throats have varying widths $w$ chosen randomly (to be detailed later), see Fig \ref{fig:square}. 
We assume that the throats connecting adjacent nodes contain the entire fluid when the porous medium is saturated. 
Hence, the nodes do not contain any fluid and serve solely as computational points for calculation of vapor and liquid pressures.


All the information about the pores and throats, including positions, lengths, states, and volumes, is stored in data arrays. For a square lattice of dimension n x m, indices $i \in [0,n-1]$ and $j \in [0,m-1]$ are associated with each node, corresponding to their position in the square lattice. Nodes are assigned  into  gas (0) or liquid (1) states, depending on the states of their neighboring throats. A node is assigned to be a gas state (0) when the first of its neighboring throats is completely emptied. Throats are also numbered depending on their positions in the lattice, with a distinction between 'vertical' and 'horizontal' throats.
Unlike nodes, throats can be empty (gas state), partially filled with liquid, or fully saturated. 
Additional arrays contain information about the horizontal and vertical throats including their widths $w_{i,j}^{\alpha}$ with $\alpha=H$ and $V$ referring to horizontal and vertical throats. The volume of a throat is given by $V_{i,j}=e*\beta a * w_{i,j}^{\alpha}$, while the instantaneous \textbf{liquid volume within the throat is represented by $V^{(l)}_{i,j}$}. Consequently, the saturation, defined as $s_{i,j} = V^{(l)}_{i,j}/V_{i,j}$, can be calculated to determine whether a throat is partially filled or fully saturated.  


\subsection{Isothermal drying algorithm}
In isothermal drying  process, gas acts as the invading phase and remains continuous, while the liquid typically fragments into numerous isolated clusters. Mass transfer is governed by vapor diffusion within the gas-filled regions of the porous space. Liquid displacement results from the competition between viscous forces and capillary pumping within the liquid-filled areas.
ness d at the interfacial boundary of the network, i.e. Fe
A typical simulation begins with a PN  fully saturated with liquid. The vapor pressure far way from the evaporation interface is assumed to be determined by the relative humidity, given as $P^{(v)}_{\infty}=RH*P^{(v)}_{sat}$, where $P^{(v)}_{sat}$ is the saturation vapor pressure of water. Vapor is allowed to escape through the top edge of network, whereas zero flux conditions are imposed on three other remaining edges. Additionally, we incorporate an  external vapor diffusion length, which describes a diffusive layer of thickness $\delta$ at the interfacial boundary of the network~\cite{plourde_pore_2003}, as detailed later in Eq.~\eqref{eq:vapor_inteface}, ensuring that  at large distances  $r\gg \delta$ the vapor pressure reaches $P^{(v)}_{\infty}$.

The procedure of the drying algorithm is described hereafter. At each step:
\begin{enumerate}
\item  \textbf{Identification of liquid clusters} \\
All liquid clusters (initially only one) are identified using the Hoshen- Kopelmann algorithm for cluster detection \cite{hoshen_percolation_1976}. Additionally, all the throats at the liquid/ gas interface of each cluster are also identified. 
    
\item \textbf{Vapor pressure field calculation} \\
For gas nodes, the vapor pressure field is obtained, using the boundary conditions in the bulk drying air ( set by the parameters $\delta$ and $RH$) and in nodes adjacent to a meniscus throat. For every gas node $i,j$ in the network, the vapor mass flux balance $\nabla \cdot \vec{J}^{(v)}=0$ is expressed as: 
\begin{equation}
    q^{(v),V}_{i,j} +  q^{(v),V}_{i+1,j} +  q^{(v),H}_{i,j-1} +  q^{(v),H}_{i,j} = 0,
    \label{eq:vaporMassBalance}
\end{equation}
where $q^{(v)}_{i,j}=J^{(v)}_{i,j} A_{i,j}$ in kg/s is the mass flow rate in the throat $i,j$ and the superscripts $^V$ and $^H$ stand for vertical and horizontal. Assuming an ideal gas behavior for the  vapor, $q^{(v)}_{i,j}$ is deduced from Fick's first law: $q^{(v)}_{i,j}=-D A_{i,j}\frac{M_v}{RT} \nabla P^{(v)}_{i,j}$ with $D$ and $M_v$ describe the vapor diffusion coefficient in the air and the molar mass of vapor. $\nabla P^{(v)}_{i,j}=\Delta P^{(v)}/(\beta a)$, with $\Delta P^{(v)}$ being the vapor partial pressure difference over the throat length and $A_{i,j} = e * w_{i,j}$ being the evaporation surface. Eq. \ref{eq:vaporMassBalance} is solved for every gas node with the finite volume method \cite{versteeg_introduction_2007}.

\item  \textbf{Liquid pressure field calculation} \\
Similarly, liquid flow within connected liquid throats (forming liquid cluster) is computed. The liquid mass flux balance  is written at every liquid node similar to in Eq. \ref{eq:vaporMassBalance}) as
\begin{equation}
    q^{(l),V}_{i,j} +  q^{(l),V}_{i+1,j} +  q^{(l),H}_{i,j-1} +  q^{(l),H}_{i,j} = 0.
    \label{liquidMassBalance}
\end{equation}
As  the flow in capillary throat is laminar, the liquid mass flow rates $q^{(l)}_{i,j}$ are given by the Hagen-Poiseuille equation: $q^{(l)}_{i,j} = -\rho_l \frac{\pi r_{i,j}^4}{8 \eta L_{i,j}} \Delta P^{(l)}_{i,j}$ where $L_{i,j}$ is the throat \textbf{liquid-filled} length, $r_{i,j}$ its radius (taken here equal to its width $w_{i,j}$ for simplicity).  The parameters $\eta$ and $\rho_l$ represent the liquid viscosity and density.  For throats at the liquid/ gas interface, there is a  competition between viscous flow and capillary force and the menisci can be either receding or stationary.  The boundary conditions are determined by  comparing the evaporation rate and liquid flow rate in them.
For throats with stationary meniscus, the liquid mass flow in the throat is equal to the evaporation rate at the meniscus: $q^{(l)}_{i,j}=q^{(v)}_{i,j}$. If the liquid cannot be supplied at the evaporation rate ({\it i.e.} $q^{(l)}_{i,j}<q^{(v)}_{i,j}$), the meniscus is defined as moving. In this case, the boundary condition on $P^{(l)}_{i,j}$ is given by the Young-Laplace law: $P^{(l)}_{i,j} = P_{atm} - 2 \gamma / w_{i,j}$, where the contact angle between liquid and throat wall is assumed to be equal to zero. Menisci are also moving in all partially filled throats and in the biggest throat of a liquid cluster. Finding which menisci are stationary and which ones are moving is done iteratively, following the method described in \cite{metzger_isothermal_2007}.

\item  \textbf{Evaluation of evaporation rates at air-liquid interfaces}\\
The evaporation rates (in kg/s), which correspond to the total mass loss from a throat at a liquid/gas interface are determined for every interfacial throats. For a throat $i,j$ (where $i \neq 0$) at the boundary of a liquid cluster, using the vapor pressure field determined in step 2, the evaporation rate is given by $q^{(v)}_{i,j}$. For every throat $0,j$ at the PN boundary, the evaporation rate is: 
\begin{equation}
q^{(v)}_{0,j} = D \frac{M_v}{RT} e w_{0,j}\frac{P^{(v)}_{sat} - P^{(v)}_{\infty}}{(1-\beta)a/2 + \delta} \label{eq:vapor_inteface}
\end{equation}
Finally, in the case of singletons $i,j$ (throats connected to two gas nodes), liquid evaporates from both menisci and the evaporation rate is given by the sum $q^{(v)}_{i,j} +q^{(v)}_{i,j+1}$ for a horizontal singleton or $q^{(v)}_{i-1,j} +q^{(v)}_{i,j}$ for a vertical singleton.

\item  \textbf{Determination of the time step} \\
The time necessary to fully empty or fill a singleton or a throat with moving meniscus selected at step 3 is determined. For a (horizontal) singleton, it is given by $t_{i,j} = |\frac{\rho_l V^{(l)}_{i,j}}{q^{(v)}_{i,j} +q^{(v)}_{i,j+1}}|$, $V^{(l)}_{i,j}$ being the volume of liquid in throat $i,j$. For a throat in a liquid cluster, the liquid flow in the throat also needs to be taken into account. 
In that case, we first obtain the net mass flow as $q^{(net)}_{i,j}=q^{(l)}_{i,j}-q^{(v)}_{i,j}$. If $q^{(net)}_{i,j}<0$, the throat is being evacuated and the time step for emptying a full or partially filled throat is given  by $t_{i,j} = |\frac{\rho_l V^{(l)}_{i,j}}{q^{(tot)}_{i,j}}|$.
If $q^{(net)}_{i,j}>0$, throat is being refilled and  the time step for refilling a partially filled throat  is determined as $t_{i,j} = |\frac{\rho_l (V_{i,j} - V^{(l)}_{i,j})}{q^{(net)}_{i,j}}|$. The minimum emptying or refilling time from all the interfacial throats is then chosen and used to define the overall time step. 

\item \textbf{Updating saturation in interfacial throats} \\
All the singletons and the interfacial throats with moving menisci selected in step 3 are assigned the mass loss or gain corresponding to the time step determined in the previous step. The throat corresponding to the smallest emptying (or refilling) time is emptied (or refilled), whereas the other throats selected are only partially emptied. 

\item \textbf{Repeat the steps 1-6} \\
The liquid phase distribution is updated, and the procedure starts again, as long as the network is not completely dry.  
\end{enumerate}

For an extensive description of the drying algorithm, one should refer to Ref. \cite{PhD_Romane}. 

\subsection{Building the pore network}
\subsubsection{Experimental micromodels}
\begin{figure}
    \centering
    \includegraphics[width=\columnwidth]{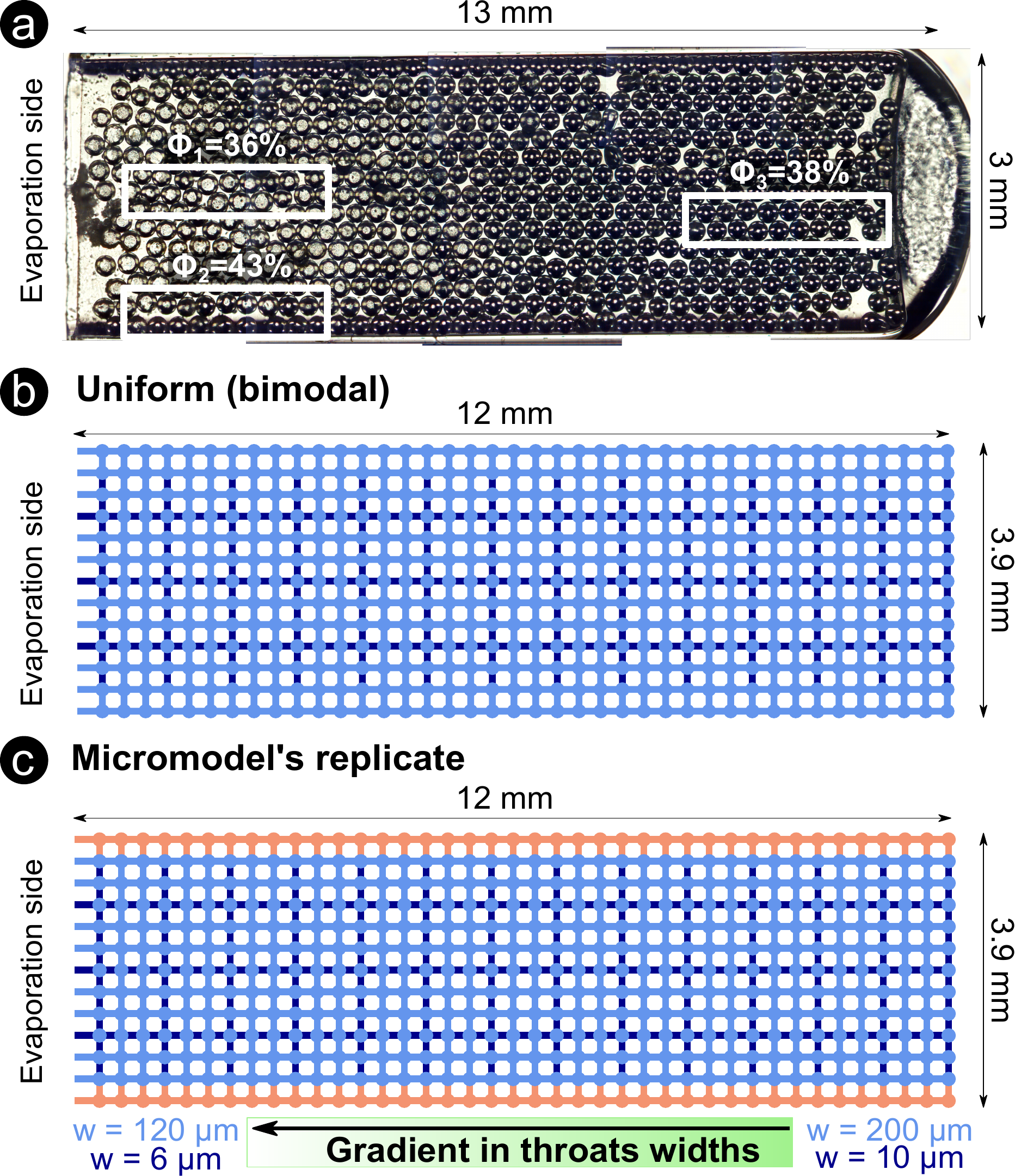}
    \caption{(a) Microscopy picture of the experimental porous media of \cite{le_dizes_castell_visualization_2024} exhibiting a heterogeneous distribution of pore size. (b) Representation of the uniform  bimodal PN w big throats of mean width of 200$\mu m$ and thin films with mean width of 8$\mu m$ are represented in light and dark blue, respectively. (c) Representation of the porous network used in the simulations to replicate the experimental micromodel of panel (a), see the text for detailed description.}
    \label{BuildingPNM}
\end{figure}

The pore network was built in order to reproduce approximately the experimental quasi-2D porous medium from Ref. \cite{le_dizes_castell_visualization_2024}, whose picture under the optical microscope is shown in Fig. \ref{BuildingPNM}(a). The porous medium was made by sintering glass beads of diameter 210 to 250 $\mu$m  in a rectangular capillary of   length   13 mm and   width is 3 mm, melted on one side. The micromodel has a thickness of roughly $h=0.2$ mm after sintering resulting in a 2D model porous medium. 
The  porosity $\phi_{exp}$ of the porous medium in experiments can be determined by image analysis:
\begin{equation}
    \phi_{exp} = 1 - \frac{1}{6} \pi d_b^{3} \frac{N}{h A} , 
\label{porosity}
\end{equation}
where $A$ is the area of the porous medium, and $N$ is the number of sintered beads of diameter $d_b=250$ $\mu$m. This leads to $\phi_{exp} \approx 42 \%$ for whole the porous medium of Fig. \ref{BuildingPNM}(a). One important matter to notice is the heterogeneities in the pore sizes and thus in the local porosity of the porous medium. Indeed, on the horizontal sides of the porous medium, there are less beads and pores are bigger. Similarly, at the evaporation side, the beads are more stacked and the pores are smaller than at the bottom of the porous medium. This is substantiated by calculating the porosity (Eq. \ref{porosity}) in the white rectangles at different locations in the porous medium, see Fig. \ref{BuildingPNM}(a).  This non-uniformity in the pore size can influence the air invasion path during the drying process. 

\subsubsection{PN choices and parameters} \label{sec:BuildingPN}

To investigate the influence of the spatial non-uniformity in the micromodels on the drying process, we constructed four  PNs with different degrees of heterogeneity. In all the pore networks, we chose $a=300$ $\mu$m, representing the distance between two nodes (estimated to be similar to the distance between two glass beads) and $\beta = 0.45$. The thickness of the network was chosen equal to $e=200$ $\mu$m, which is roughly the thickness of the experimental micromodel after sintering. The length of every throat in all the PNs is given by $\beta a = 135$ $\mu$m. However, the width of the throats of the four PNs were varied differently, while keeping the same mean width of 200 $\mu$m. The PNs consist of 13x40 nodes, containing 1000 throats, resulting in a total length of 12 mm and a width of 3.9 mm. Authors of Ref. \cite{yiotis_effect_2003, laurindo_numerical_1998} showed that the presence of thin films cannot be neglected and has a strong influence on the drying patterns in PN models. To account for the presence of thin films in the experimental micromodels, we added thinner throats in the PNs with mean width of 8 $\mu$m (see throats in deep blue in Fig. \ref{BuildingPNM}(b)). More specific details about each of the four micromodels is provided below.   

\begin{enumerate}
    \item [$M1.$]\textbf{Uniform bimodal PN:} A PN with a uniform throat width distribution was built to serve as reference. In this pore network represented in Fig. \ref{BuildingPNM}(b), the big vertical and horizontal throats (light blue) widths are chosen randomly from a uniform distribution in the interval $[140,260]$ $\mu$m with a mean width of $200$ $\mu$m. The narrow throats widths (deep blue in Fig. \ref{BuildingPNM}(b)) are chosen randomly from a uniform distribution in the interval $[6,10]$ $\mu$m.
    \item [$M2.$] \textbf{PN with a horizontal gradient in the throats widths:} To mimic the smaller pores on the evaporation side observed in experiments, a uniform throat width gradient was imposed along the length of the PN. The wide throats of the closed side of the PN ($i=n-1)$ have widths of $260 \pm 10$ $\mu$m, whereas the ones at the evaporation side  ($i=0)$ have a width  of $140 \pm 5$ $\mu$m. For the narrow throats, a similar gradient is imposed with  a width  of $10 \pm 0.2$ $\mu$m for $i=n-1$ and $6 \pm 0.2$ $\mu$m for $i=0$. 
    \item[$M3$.] \textbf{ PN with vertical gradients in the throats widths:} The influence of having wider throats on the lateral sides was also investigated by defining a PN with gradients along the width of the porous network. The narrower throats in the center of the porous network $j = 6$ have a width of $120 \pm 5$ $\mu$m, whereas the ones on the sides ($j=0$ and $j=12$) have a width of $240 \pm 10$ $\mu$m. In the narrow throats, for $j=6$, $w= 6 \pm 0.2$ $\mu$m and, for $j=0$ and $j=12$, $w = 10 \pm 0.2$ $\mu$m. 
    \item[$M4.$] \textbf{Replicate of the micromodel PN:} Finally, a PN, sketched in Fig. \ref{BuildingPNM}(c), was constructed to reproduce the experimental micromodel, featuring a gradient in throat width along the length of the porous medium and wider throats at the edges, shown in orange in Fig. \ref{BuildingPNM}(c). The orange throats at the edges have widths comprised between 225 to 230 $\mu$m. As depicted in Fig. \ref{BuildingPNM}(c), the throats on the closed side of the PN have a width  of $194 \pm 3$ $\mu$m, while those  at the evaporation side have  widths of $122 \pm 3$ $\mu$m. The narrow throats, shown in dark blue in Fig. \ref{BuildingPNM}(c), have widths ranging from $10 \pm 0.2$ $\mu$m in the inner part of the PN to $6 \pm 0.2$ $\mu$m at the evaporation side. 
    \end{enumerate}
\begin{figure*}[http]
    \centering
    \includegraphics[width=0.99\textwidth]{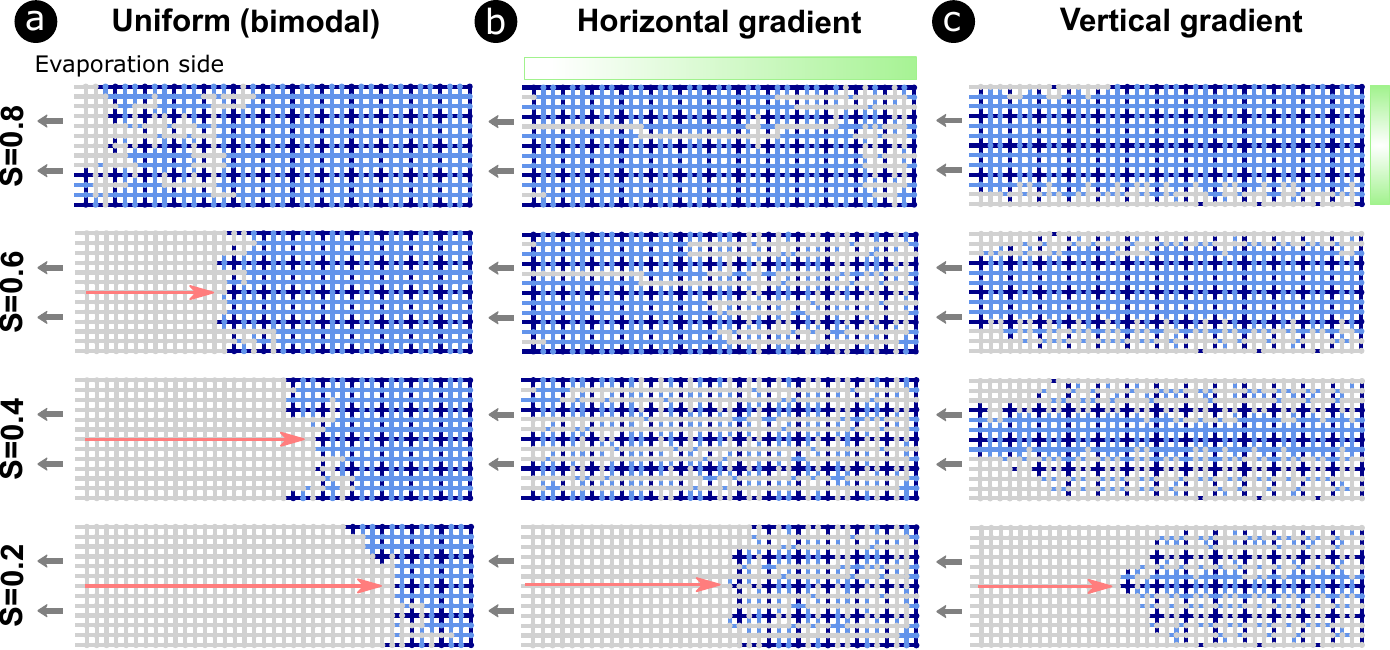}
    \caption{Snapshots of different stages of the drying of water quantified by the overall saturation $S$ in (a) the uniform bimodal PN, (b) in the PN with a horizontal gradient in the throats widths  and (c) in the PN with vertical gradients in the throats widths (c). The liquid filled throats are depicted in blue (deep blue for throats with $w\leq10$ $\mu$m) and the empty throats are shown in light grey. The green bands in b and c illustrate the gradients in the width of throats where the wider throats are encoded with the green color and the narrower ones at the white color}
    \label{Heterogeneities}
\end{figure*}

\begin{figure}[http]
    \centering
    \includegraphics[width=0.9\columnwidth]{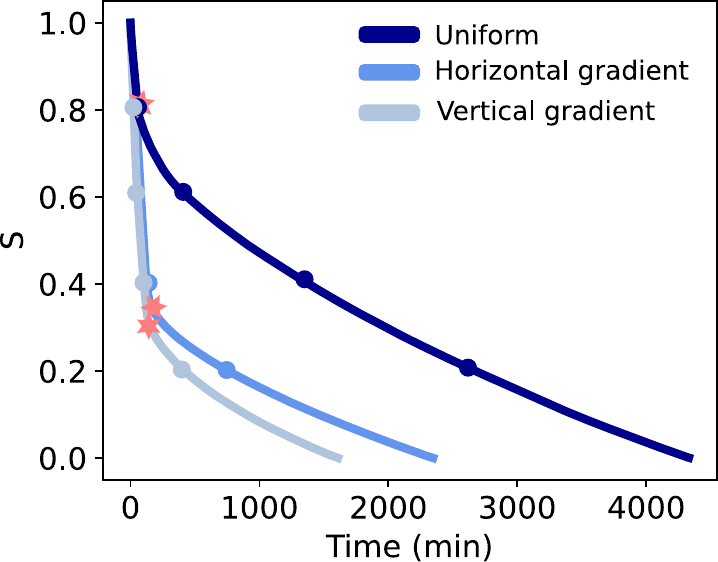}
    \caption{The overall saturation of liquid $S$ as function of time in the three PN models shown in  panels (a-c) of Fig.~\ref{Heterogeneities} . The disk symbols on the curves indicate the times corresponding to the different snapshots in panels (a-c) of Fig.~\ref{Heterogeneities} and the pink stars indicate  the transition from the CRP to the FRP in the three PNs.}
    \label{Heterogeneities_DC}
\end{figure}

We  emphasize that the porous networks used in the simulations in some aspects differ   from those used in the experiments, as the network of pores and throats are respectively 2D and quasi-2D, respectively. Additionally, spatial partitioning of our square lattice PN differs from  irregular  porous network of sintered beads. Therefore, our modeling accounts only for the key characteristics of the micromodel, such as the overall pore volume, mean porosity, and spatial gradients in pore size.
  For all constructed PNs, the mean porosity is $\Phi = 39\%$, which   closely approximates the porosity of  experimental system $\Phi_{exp} = 43\%$.  Across all networks, the total throat volume is $V_{pores} = 3.5$ mm$^3$. We define the overall saturation as 
  \begin{equation}
      S = \frac{m(t)}{m_0},
  \end{equation}
   where $m(t)$ is the liquid's instantaneous mass and $m_0$ is the mass of saturated porous medium at the outset.

In the following simulations, we assume that a liquid with the same properties as water is evaporating. The water vapor diffusion coefficient in air is set to $D = 2.4*10^{-5}$ m$^2$.s$^{-1}$, the surface tension of water to $\gamma=72$ mN.m$^{-1}$, and the liquid density to $\rho_{l} = 1000$ kg.m$^{3}$. Unless stated otherwise, the fluid viscosity is $\eta =0.001$ Pa.s. The evaporation conditions are fixed with a relative humidity of $RH = 45 \%$, a vapor saturation pressure of $P^{(v)}_{sat} = 2340$ Pa, and  a temperature of $T=20^{\circ}$C. Finally, the diffusive length above the porous medium is set to $\delta=50$ $\mu$m.

\section{Effects of spatial heterogeneity of pore size on drying of water} \label{sec:WaterPNM}

\subsection{Effect of the pore size distribution} \label{sec:HetPN}

Here, we study the effect of the heterogeneity in the pore size distribution on the drying of water in the porous media and the corresponding drying regimes. We compare PNs with uniform throat width distribution to those with a horizontal or vertical gradients in throats widths described as models $M1-M3$ in section \ref{sec:BuildingPN}. Snapshots of the drying process at identical overall saturation levels are shown for all three cases in Fig. \ref{Heterogeneities}. The snapshots in Fig.~\ref{Heterogeneities} clearly demonstrate that the air invasion path strongly depends on the throat size distribution. In the uniform bimodal PN (Fig. \ref{Heterogeneities}(a)), the main (largest) liquid cluster disconnects from the evaporation side at early drying stages when $S \le 0.8$.  In contrast, for networks with horizontal and vertical gradients (Figs. \ref{Heterogeneities}(b) and (c)), the largest liquid cluster stays connected to the evaporation side until late  stages of drying, only disconnecting when $S < 0.4$.\\

In the horizontal gradient network, see Fig. \ref{Heterogeneities}(b), the invading gas forms a nearly straight path from the evaporation side to the sealed side, splitting the network into two liquid clusters, both connected to the evaporation side. The wider throats at the bottom desaturate first, while narrower ones near the evaporation side stay filled with liquid. With no flow between the clusters, they evaporate independently, at rates determined by the number of throats on the evaporation side. Once all the throats connected to the evaporation side empty ($S \approx 0.35$), the remaining liquid is mostly in narrow throats (deep blue), and a dry front forms that advances inward.\\

In the vertical gradient network (Fig. \ref{Heterogeneities}(c)), air first invades the wider throats at the lateral sides, progressing toward the closed end of the porous medium. Desaturation moves from the wider throats (top and bottom) to the narrower ones (middle), forming two horizontal 'partial desaturation' fronts.  However, the largest liquid cluster, containing  narrower throats concentrated in the center, stays connected to the evaporation side until $S \approx 0.35$, when the two fronts converge and a dry front forms.

\begin{figure*}[http]
    \centering
    \includegraphics[width=0.99\textwidth]{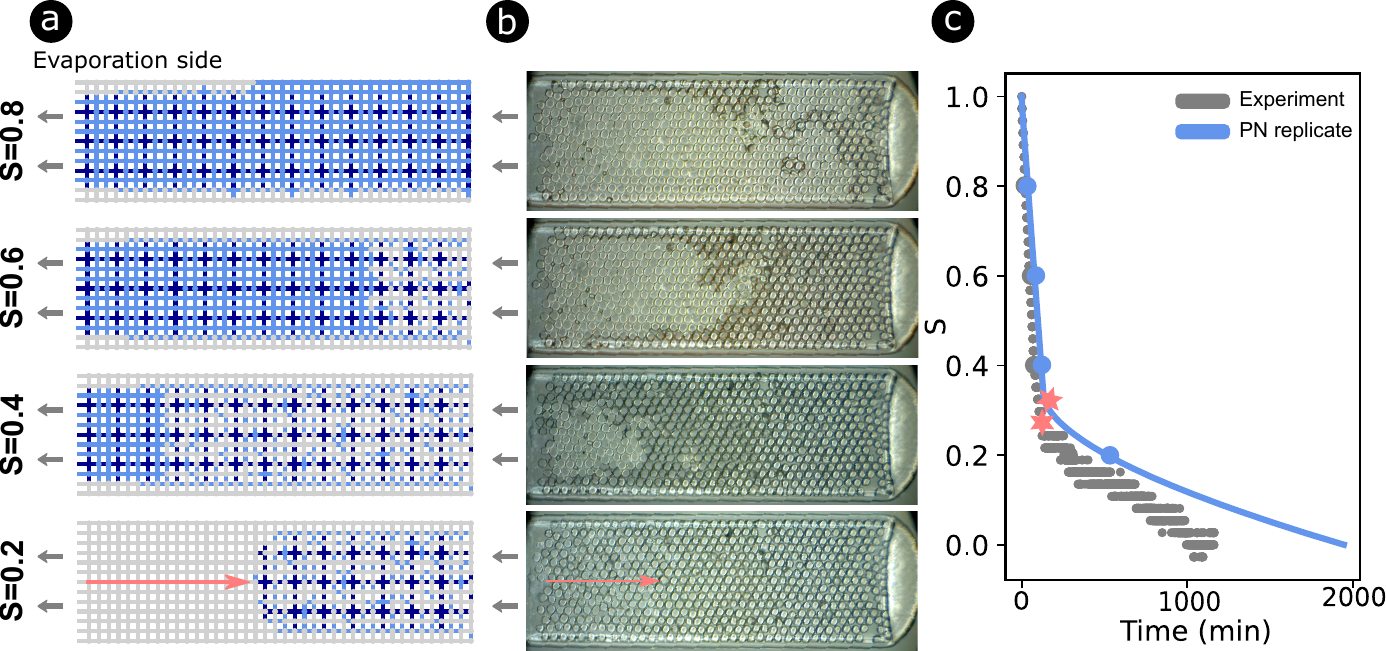}
    \caption{(a) Snapshots at different moments of the drying of the solvent in the PN constructed based on experimental quasi-2D medium compared to the experimental results shown in panel (b). (c) Saturation level $S$ of liquid in the PNM and in the experiments as functions of time. The big dots correspond to the saturation levels shown in panels (a) and (b).}
    \label{fig:solvent}
\end{figure*}

The overall saturation $S$ of networks over time is presented in Fig. \ref{Heterogeneities_DC}. For all the PNs, two main drying regimes, characteristic of the drying in porous media, can be identified: the constant rate period (CRP) and the falling rate period (FRP)~\cite{coussot_scaling_2000, prat_recent_2002}. The CRP is characterised by a fast and almost constant drying rate. At that stage, the liquid is connected to the evaporation surface where it evaporates. Once the liquid is disconnected from the evaporation surface, the FRP starts, characterised by the development of a dry front receding in the porous network, marked by pink arrows in Fig. \ref{Heterogeneities}(a)-(c), leading to an abrupt decrease in the evaporation rate. The onset of FRP is indicated by pink stars in Fig. \ref{Heterogeneities_DC}.   This agrees with the drying patterns observed in Fig.~\ref{Heterogeneities}(a)-(c). The relative duration of the CRP and FRP varies significantly with changes in throat width distribution across the three networks. In the uniform bimodal network (Fig. \ref{Heterogeneities}(a)), the CRP is brief, with the FRP beginning at a high saturation of $S = 0.8$, resulting in  a prolonged duration for complete drying. In contrast, in the gradient networks (Figs. \ref{Heterogeneities}(b) and (c)), liquid remains connected to the evaporation surface for most of the drying process, with significant desaturation occuring during the CRP, up until $S \approx 0.4$, leading to a shorter overall drying time. For all  networks, the narrow throats ($w \leq 10$ $\mu$m, deep blue) remain fully saturated during the CRP and empty only during the FRP. The influence of the very narrow throats, mimicking effects of thin films, is investigated more in-depth in Appendix \ref{sec:Appendix4_TT}.

\begin{figure*}[http]
    \centering
    \includegraphics[width=0.99\textwidth]{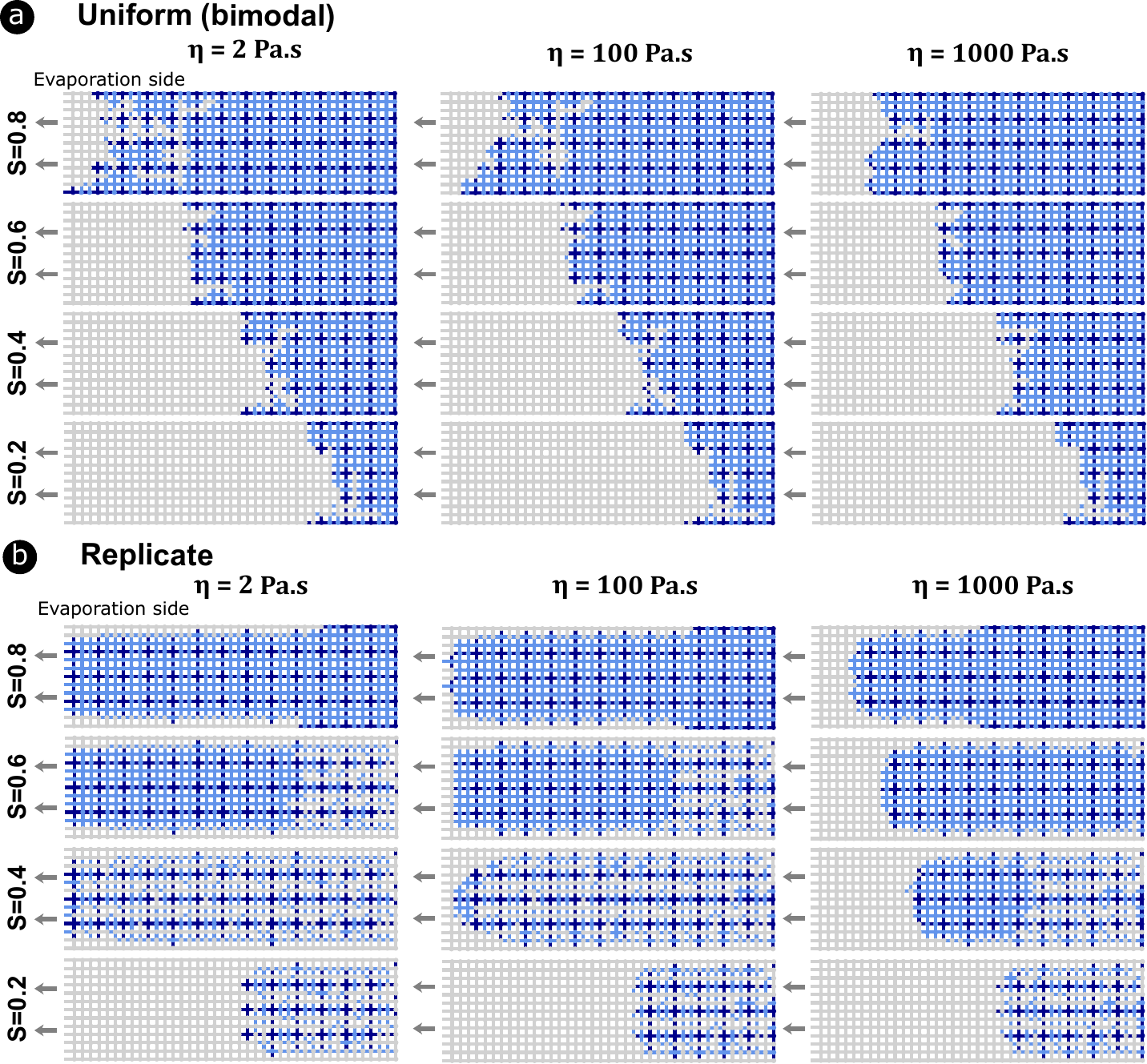}
    \caption{(a) Snapshots at different moments of the drying of a pure liquid with $\eta=2$ Pa.s, $\eta=100$ Pa.s and $\eta=1000$ Pa.s in uniform (bimodal) PN and (b) in the replicate PN. The color scheme is similar to the one used in Fig. \ref{Heterogeneities}.}
    \label{fig:uniformVisco}
\end{figure*}

\subsection{Validation of replicate PN with water experiments}
As heterogeneity in the pore size distribution strongly influence the drying patterns and kinetics in our 2D models, we now study the drying in the PN model M4 of section \ref{sec:BuildingPN}, which combines heterogeneity in the vertical and horizontal directions to replicate the experimental micromodel of Ref. \cite{le_dizes_castell_visualization_2024}.  In Ref. \cite{le_dizes_castell_visualization_2024}, the solvent that evaporates is a mixture of ethanol and water. The drying of binary mixtures in PNM has previously been studied in Ref. \cite{freitas_pore_2000} and it requires to implement the selective evaporation of both liquids. In the following, for simplicity, we assume the solvent is only water. The transport properties of water and ethanol are similar in the liquid phase, so they minimally affect fluid flow compared to the sol-gel process. Although ethanol is more volatile in the gas phase, it does not impact the overall process. The vapor diffusion length $\delta$ at the network boundary was set to $\delta = 50$ $\mu$m, based on experimental data \cite{le_dizes_castell_visualization_2024}, to match the initial evaporation rate in the PNM.

In the replicated PN, shown in Fig. \ref{fig:solvent}(a), air invades first the widest throats located along the top and bottom edges of the network, as seen in the first  snapshot of Fig. \ref{fig:solvent}(a). Once the wide throats on the lateral sides are nearly dried, the larger throats on the closed side of the network begin to empty, while those near the evaporation side remain filled.  This process is illustrated in the second and third snapshots of Fig. \ref{fig:solvent}(a). A \emph{partial desaturation front} forms, which progresses from the widest throats on the right side to the narrowest ones at the evaporation side. This process carries on until most of the throats are empty. The air invasion pattern in the PN is similar to the experimental results presented in Fig. \ref{fig:solvent}(b). It is a combination of the air invasion paths in Fig. \ref{Heterogeneities}(b) and (c). Once most of the throats have become empty in the network, see the third snapshots of Fig. \ref{fig:solvent}(a) and (b), the FRP begins, and the last and narrowest throats empty in both PN and experiments as shown in final snapshots of Fig. \ref{fig:solvent}(a) and (b). The pink arrows indicate the position of the receding drying front. The drying curve in the replicated PN also agrees with the experimental results, as presented in Fig. \ref{fig:solvent}(c). In conclusion, our replicated PN model   captures well both air invasion paths and drying kinetics, even though we only considered water evaporation instead of the solvent mixture.

\section{Effect of viscosity on drying of pure liquids}
\label{sec:PNM_Visco}
\begin{figure}[http]
    \centering
    \includegraphics[width=0.9\columnwidth]{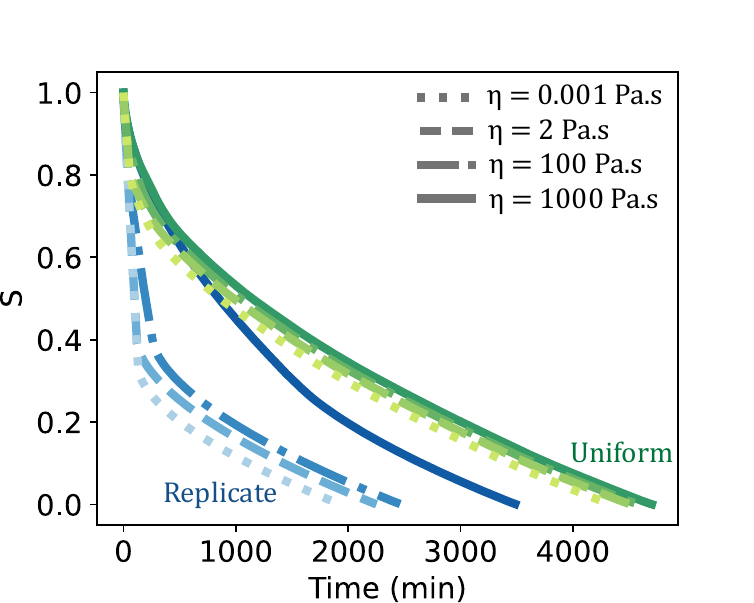}
    \caption{The overall saturation $S$ vs time  for different fluid viscosity values in the uniform PN (green) and in the replicated PN (blue).}
    \label{fig:DCuniformVisco}
\end{figure}

As described extensively in Ref. \cite{le_dizes_castell_visualization_2024}, the sol-gel transition is accompanied by a sharp increase of viscosity from $\eta \approx 10^{-3}$ Pa.s to $\eta \approx 80$ Pa.s at the gel point. We investigate the effects of a constant and uniform viscosity on the drying process in the uniform (bimodal) and replicated PNs. The drying patterns for each PN at three viscosity values ($\eta = 2$, 100, and 1000 Pa·s) are shown in Fig. \ref{fig:uniformVisco}. For the network with uniform throat width distribution (see \ref{fig:uniformVisco}(a)), increasing viscosity by nearly three orders of magnitude has a very weak impact on the air invasion path and the drying kinetics as shown in Fig. ~\ref{fig:DCuniformVisco}. This is because, even at lowest viscosity, a dry receding front advances early in the drying process, and the CRP is very brief. 

In contrast, for the replicated PN shown in Fig. \ref{fig:uniformVisco}(b), where the CRP lasts a significant portion of the drying process at $\eta = 0.001$ Pa·s, viscosity has a noticeable impact on both the air invasion path and drying curves for viscosities of $100$ and $1000$Pa.s. Indeed, at $\eta = 100$ Pa·s, the CRP is shorter compared to $\eta = 0.001$ Pa·s or $\eta = 2$ Pa·s. Higher viscosity also reduces the roughness of the drying fronts.  However, it is only at a much higher viscosity ($\eta = 1000$ Pa·s) that viscous losses become too significant to replace the evaporating liquid at the surface, eliminating the constant rate period and causing the drying kinetics to resemble those of the uniform PNs, as seen in Fig. ~\ref{fig:DCuniformVisco}.

Thus, a heterogeneous PN and a very high viscosity of the order of $\eta=1000$ Pa.s is necessary to observe a significant impact on the drying process. We emphasize these results are valid for our chosen types of PN which include both narrow and wide throats. Authors of Ref. \cite{metzger_isothermal_2007} showed that for larger networks with narrower pore size distribution, the viscosity effects can be significant even for $\eta=0.001$ Pa.s leading to a shorter CRP. 

\section{Sol-gel transition with evaporation} 
\label{sec:SolGelPNM}
\subsection{Implementing the heterogeneous increase of viscosity based on experimental results} \label{sec:viscoImpl}
\begin{figure}[http]
    \centering
    \includegraphics[width=\columnwidth]{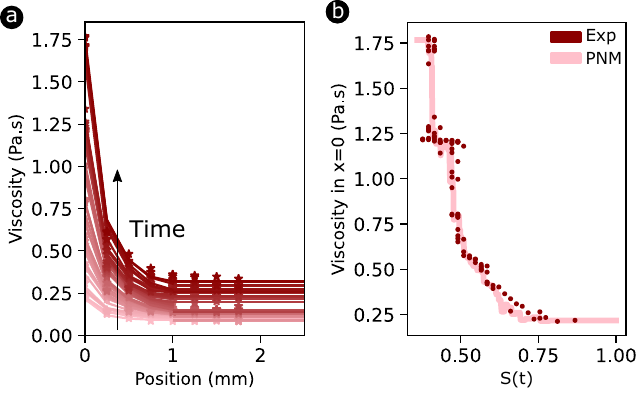}
    \caption{(a) Viscosity within experimental porous medium depicted in Fig.~\ref{fig:experimentalResults} as a function of distance from entrance of porous medium measured every 30 min until $t=2940$ min using Fluorescence lifetime imaging microscopy \cite{le_dizes_castell_visualization_2024}. (b) Viscosity at the evaporation side of the porous medium $x=0$ as a function of the saturation.}
    \label{viscoIncrease}
\end{figure}

In the quasi-2D porous medium shown in Fig. \ref{fig:experimentalResults}(a), the increase in the liquid viscosity during the sol-gel transition depends on the distance from the evaporation side. Interestingly, over time a gradient in viscosity develops within the porous medium. Fig. \ref{viscoIncrease}(a) shows the increase in viscosity over a time span of $1<t<2940$ min, within the first 2.5 mm of the porous medium from the evaporation side. The increase in viscosity is strongest at the entrance of the 2D porous medium, where its value increases from the that of solvent to about 1.75 Pa.s at the latest measurement time $t=2940$ min.  An overall increase in viscosity is nonetheless observed everywhere across the porous medium. As shown in the previous section, a uniformly high viscosity of 2 Pa·s has negligible impact on  drying kinetics in porous media. We  thus now examine the effect of a viscosity gradient, similar to Fig. \ref{viscoIncrease}(a), during gel formation on capillary liquid flows across the pores and drying kinetics. ub
\begin{figure*}[http]
    \centering
    \includegraphics[width=0.99\textwidth]{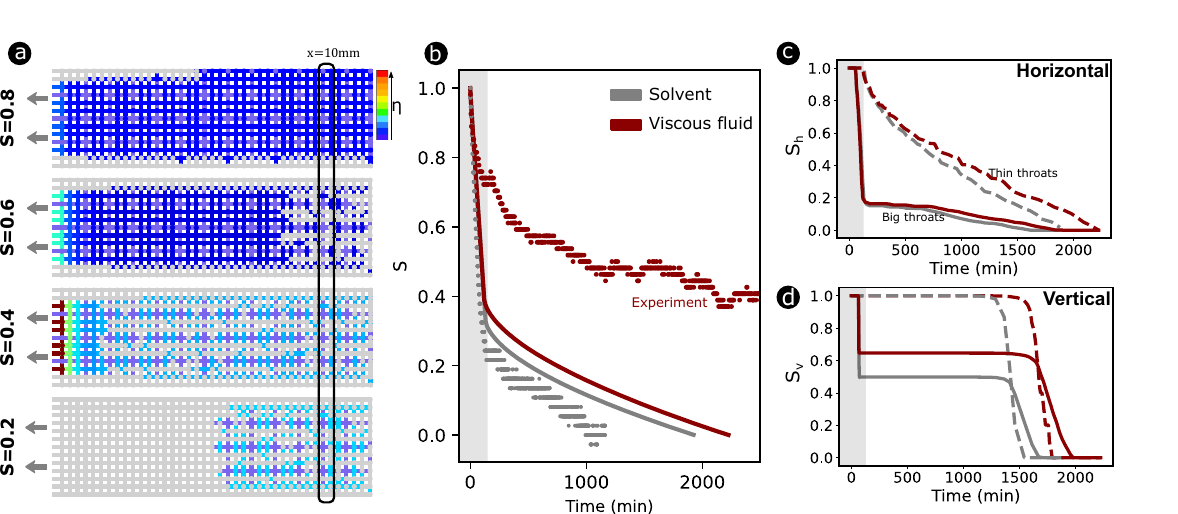}
    \caption{(a) Snapshots of drying of fluid with viscosity gradients in replicate PN where the purple throats denote the narrow throats with $w\leq10$ $\mu$m. b) Evolution of saturation $S$ over time for the solvent and the fluid with the spatio-temporal viscosity profile extracted from experiments shown in Fig.~\ref{fig:experimentalResults}. The shaded area denotes the CRP. This is compared to the experimental results for the solvent (grey dots) and for the fluid undergoing sol-gel transition (red dots). (c) and (d). The mean (local) saturation in narrow and wide throats in the directions parallel (horizontal) and perpendicular (vertical) to the mean direction of air invasion flux at $x=10$ mm (see the position in a) plotted as a function of time for the solvent (grey) and the fluid with viscosity gradient (red).}
    \label{fig:InfluenceVisco}
\end{figure*}
To explore the influence of spatio-temporal viscosity changes on evaporation, we implement a time-evolving viscosity in the PN model, focusing on how these variations affect the air invasion path in the replicate PN. To implement the viscosity gradient profile in the replicate PN model, we determined from the experimental data the evolution of the viscosity as a function of the overall saturation $S$ of the micromodel $\eta(S(t),x)$. Then, at every time step in the PN model, the saturation $S(t)$ is determined and the viscosity of the fluid is updated in order to be equal to the experimental viscosity at the same saturation level. Fig. \ref{viscoIncrease}(b)  illustrates the viscosity at position $x=0$ (evaporation side) as a function of the saturation level $S(t)$ in both the 2D porous medium and the PNM. This ensures that the viscosity profiles in the PNM match those observed in the experiments at 
{\it identical saturation levels}.

\subsection{Influence of viscosity gradient on drying kinetics}

Fig. \ref{fig:InfluenceVisco}(a) shows the resulting air invasion path of a liquid with a viscosity gradient profile similar to experimental data of Fig. \ref{viscoIncrease} at different saturation levels. The observed air invasion patterns in the replicated PN are similar to those in the experimental micromodel up to $S=0.4$ (see for comparison Fig. \ref{fig:experimentalResults}(a)). The viscosity increases everywhere during the drying, with a sharper increase at the evaporation side, where it reaches values up to $\approx 2$ Pa.s. The increase in viscosity is not sufficient to strongly influence the drying pattern in comparison to the drying pattern of the solvent shown in Fig.~\ref{fig:solvent} and the drying rate as demonstrated in Fig. \ref{fig:InfluenceVisco}(b). 
In the PNM, for $S<0.25$, a dry receding front develops from the evaporation side, which penetrates inside the PN, until the whole fluid evaporates. Although the air invasion path in the PN closely resembles to the experimental results during the sol-gel transition, as demonstrated in Fig. \ref{fig:InfluenceVisco}(b) the drying rates of PN with similar viscosity profile is substantially faster than the experimental counterpart.

Similar to  fluids with high and uniform viscosity, when the FRP begins, more liquid remains trapped in narrow throats. This is also observed experimentally, as gel forms thin films everywhere inside the porous medium \cite{PhD_Romane}. The trapping of liquid in the PNM is thus further explored in Fig.~\ref{fig:InfluenceVisco}(c) and (d), where partial saturation levels for the viscous fluid and solvent in horizontal and vertical directions are compared. These figures show saturation levels of horizontal $S_H$ and vertical $S_V$ throats as functions of time, while discerning wide (thick lines) and narrow  (dotted lines) throats. For both kinds of fluid, we observe a significant difference between drying of vertical and horizontal throats. Wide vertical throats retain higher saturation level than their horizontal counterparts throughout drying. This difference can be attributed to the asymmetric design of the porous medium, favoring horizontal air invasion, resulting in a horizontal viscosity gradient. The viscosity gradient has a minor effect  on horizontal throats saturation but significantly impacts vertical throats as visible in Fig.~\ref{fig:InfluenceVisco}(d). 
At $t \approx 200$ min, the first \emph{desaturation} front reaches $x=10$ mm, emptying a large fraction of wide throats, while narrower ones remain almost saturated. For the  fluid with imposed viscosity gradient, the saturation in wide vertical throats is nearly 20\% higher than for the solvent. Narrow vertical throats  remain fully saturated during the CRP, whereas their wide counterparts desaturate partially during CRP and retain their saturation value until $t\approx 2000$ min up to overall saturation level  $S\leq0.25$. 

Our PNM with an imposed viscosity gradient extracted from experiments captures some key features of drying during the sol-gel transition, particularly the air invasion path and thin-film trapping, aligning well with the experimental results of Ref. \cite{le_dizes_castell_visualization_2024}. However, some aspects of  experiments reported in Ref. \cite{faiyas_transport_2017}, especially skin formation  are not taken into account in our modeling approach. In drying polymer solutions or colloidal suspensions, particles are advected with the liquid and deposit at the evaporation side, leaving throats near the entrance of porous medium partially saturated (in experiments, saturation remains around 0.25 at the end). Our model does not account for polymer displacement, and it also fails to capture the significant drop of evaporation rate  during the sol-gel transition. In the next section, we propose a revision of the PNM to rationalize the reduced evaporation rate due to skin formation.

\subsection{Accounting for skin formation  with viscosity-dependent vapor pressure}
As discussed in the introduction, for coating-forming silanes drying can lead to the formation of a skin at the free surface (a process also called ''skinning') which reduces or prevents solvent evaporation afterward \cite{brinker_review_1992, brinker_sol-gel_2013, cairncross_predicting_1996}. To account for the effect of skin formation in our PNM, we use the experimental insights obtained from the study of sol-gel transition in round capillary in Ref. \cite{le_dizes_castell_visualization_2024} where skin formation at the evaporating meniscus was observed. A brief recap of these results are reproduced in Fig. \ref{fig:roundcap} of Appendix \ref{sec:AppRoundCap}. The drying in a round capillary happens in two main regimes: a first period during which the fluid dries at the same rate as the pure solvent and a second period with strongly reduced evaporation rate beginning after the formation of a viscous layer of viscosity  $\eta_c \approx 0.22 \pm 0.03$ Pa.s at the evaporation surface. For the second drying regime, the reduced evaporation rate can be deduced from experimental data of Ref. \cite{le_dizes_castell_visualization_2024} as described in more details in the Appendix. The drying rate can be approximated by considering that the vapor pressure above the meniscus is equal to $P^{(v)}_c = \Delta P^{(v)} + P^{(v)}_{\infty}$ where $\Delta P^{(v)} = 40$ Pa (see Appendix). Because round capillaries can, to some extent, be considered as model pores, we expect the skin formation to happen everywhere at the pore scale during the drying of the porous medium. 

To investigate the influence of the skin formation in the porous medium, additional conditions are implemented in the PNM, based on the experimental insights from the evaporation in the round capillary. At every time step and in every throats at the boundary of a liquid cluster, the vapor pressure above the meniscus for water evaporating can be approximated as: 
\begin{equation}
\left\{
    \begin{array}{ll}
        P^{(v)} =P^{(v)}_{sat} = 2340 \mbox{ Pa} & \mbox{if } \eta \ll \eta_c \\
        P^{(v)} = 40+ P^{(v)}_{\infty} \approx 1100 \mbox{ Pa}& \mbox{if } \eta \gg \eta_c 
    \end{array}
\right.
\end{equation}
as $P^{(v)}_{\infty}=RH*P^{(v)}_{sat}= 1053$. To avoid discontinuities in vapor pressures, the following sigmoid-based smoothing is implemented: 
\begin{equation}
P^{(v)}(\eta) = P^{(v)}_{sat} + \frac{P^{(v)}_c – P^{(v)}_{sat}}{1+\text{exp}(-100(\eta-\eta_c))}
\label{eq:pv=f(n)}
\end{equation}
where $\eta_c = 0.22$ Pa.s and $P^{(v)}_c = 1100$ Pa.s. This function reproduces the very sharp decrease in the vapor pressure observed experimentally for $\eta_c = 0.22$ Pa.s and Eq. \eqref{eq:pv=f(n)} is implemented in every throats to model the vapor pressure dependency on the local viscosity.

\begin{figure}[http]
    \centering
    \includegraphics[width=0.9\columnwidth]{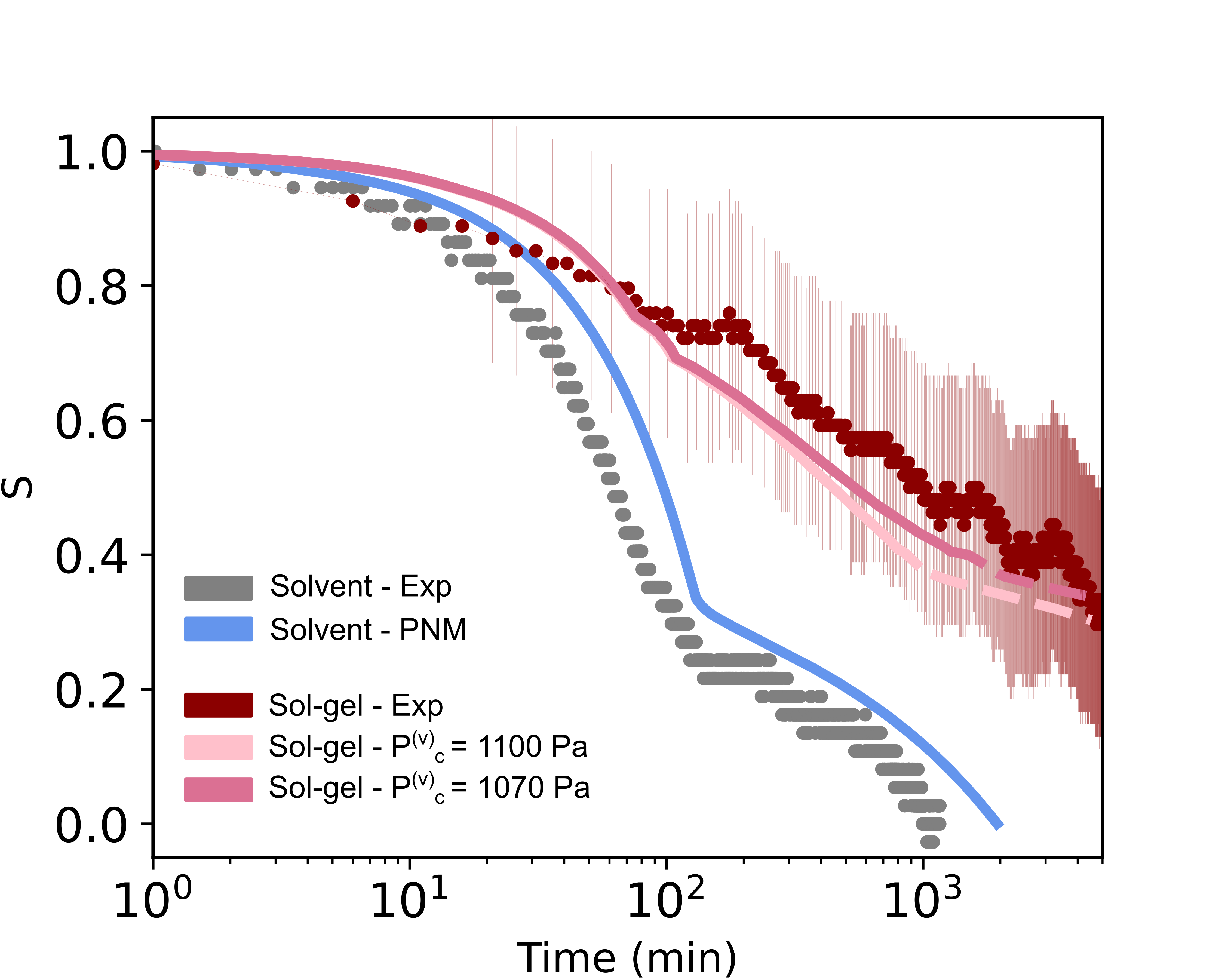}
    \caption{The overall saturation level as function of time for the drying of the solvent and during the sol-gel transition compared to the predictions of replicate PNM when accounting for the dependence of vapor pressure on viscosity.}
    \label{DC_PNM}
\end{figure}

Fig. \ref{DC_PNM} shows the drying kinetics obtained in replicate PN for the solvent and the fluid with viscosity gradient profile,  incorporating a viscosity-dependent vapor pressure described by Eq.~\eqref{eq:pv=f(n)}. The latter  accounts implicitly for the skin formation effects within throats containing sufficiently viscous fluids. For comparison,  the experimental results of drying rates of solvent and fluid undergoing sol-gel transition are also shown.
A good agreement between the PNM and experimental results for the fluid undergoing sol-gel transition is observed when the viscosity cut-off is implemented with $\eta_c = 0.22$ Pa.s and for a decreased vapor pressure of $P^{(v)}_{c} = 1100$ Pa above the throats where $\eta > \eta_c$. For $P^{(v)}_{c} = 1070$ Pa, the drying kinetics are accurately captured. The discrepancy with the experimental results of the round capillary is small considering that the measurement in the round capillary were performed at $RH=35 \%$ (against $RH = 45 \%$ in the porous medium) and that several assumptions were made when calculating $P^{(v)}_{c}$ (constant evaporation rate and only water evaporating). 

The proposed dependence of vapor pressure  on viscosity given by Eq.~\eqref{eq:pv=f(n)} at the single-throat level effectively predicts the drying kinetics of the entire porous medium. Compared to the solvent, the drying rate decreases strongly from the beginning of the drying and it continues to decline over time. At $t \approx 6000$ min (when the experiment was stopped), the saturation of the PN remains above 0.25, closely matching the experimental results.
Thus, our model can capture both the air invasion path and drying kinetics.  We note that the stage 4 of Fig. \ref{fig:InfluenceVisco}(a) is not reached within experimental timeframe studied in reference~\cite{le_dizes_castell_visualization_2024}. The good agreement with experimental results suggests that the local skin formation in the throats governs the drying process of the whole porous network. A  solid-like 'gel' front forms at the interface between the low viscosity region and the gelled region, which  gradually advances into the porous medium. \\

\section{Concluding remarks} \label{sec:conclusion}

In this work, we have developed a novel pore network model (PNM) customized for experiments on sol-gel transitions in quasi-2D porous media. Our pore network replicates  the main features of the experimental system of Ref. \cite{le_dizes_castell_visualization_2024}, {\it i.e.}, the mean porosity and, spatial gradients in the pore size distribution. 
To elucidate the role of pore size distribution, we first conducted simulations of pure liquid evaporation in PNs with different profiles of throat-width distributions. From simulations of  water evaporation, we found that the air invasion path and the drying rate are primarily dictated by the pore size distribution. Notably, the outset of FRP is concomitant with the time that largest liquid cluster becomes disconnected from the evaporation side.

In a uniform PN, where the wide throats are randomly located and not connected to each other, the main liquid cluster becomes quickly disconnected from the evaporation side resulting in early commencement of FRP. Introducing a spatial gradient in throat width, either vertically or horizontally  provided that the smallest throats are positioned at the evaporation side, allows 
 for a controlled air invasion in the PN and   extends the duration of CRP, as observed experimentally and recently reported in Ref. \cite{fei_pore-scale_2024}. Furthermore, a PN incorporating a small fraction of randomly distributed   very narrow throats ($w \leq 10 ,\mu$m) better accounts for liquid thin film formation, offering a more accurate representation of pure liquid drying and better agreement with experimental drying curves.

To understand the effect of viscosity increase occurring during sol-gel transition, we first examined the impact of pure liquid viscosity  on drying kinetics in both uniform and replicate PNMs. In uniform bimodal PNs, even a $10^6$-fold increase in viscosity had no significant effect on drying patterns or kinetics. However, in replicate PNs with spatial throat size gradients where narrower throats are located at the evaporation side, a viscosity of around 1000 Pa·s was necessary to noticeably affect the drying process. At high viscosity values, more liquid remains trapped in large pores, slowing down the drying process and producing smoother evaporation fronts, characteristic of viscous fingering.

Next, to capture the heterogeneous  viscosity increase during the sol-gel transition in porous media, we integrated the experimentally measured spatio-temporal viscosity profile  from  Ref. \cite{le_dizes_castell_visualization_2024} in our replicate PNM. Next, to capture the heterogeneous viscosity increase during the sol-gel transition, we integrated the experimentally measured spatiotemporal viscosity profile from Ref. \cite{le_dizes_castell_visualization_2024} into our PNM. Our findings confirm that the air invasion path remains primarily governed by the pore size distribution. Even though viscosity increases during the sol-gel transition, reaching a maximum of 80 Pa·s, it remains insufficient to significantly alter liquid flow in the porous medium, leaving the air invasion path largely unchanged.

Finally, we investigated how local skin formation at throat menisci influences the overall evaporation rate in porous media. Experimentally, we identified a simple relationship between meniscus viscosity and vapor pressure during the sol-gel process, which can be approximated as a reduction in vapor pressure once viscosity exceeds a critical threshold. Generalizing this effect from a single pore to the throats within an entire porous network, we demonstrated that it can rationalize the drying kinetics of complex fluid undergoing sol-gel transition during evaporation. The overall slow evaporation in the porous medium can be attributed to reduced evaporation rates at throats where menisci have developed a skin. In other words, the localized skin formation at the throat level  dictates the drying kinetics of the entire network.

 This work represents a first step toward understanding and predicting the drying kinetics of complex fluids undergoing phase changes in porous media, with broad applications in areas such as mortar and concrete formation, microbial activity in soils, and the consolidation of decaying stones. However, further research is needed to extend this model to account for particle transport during drying, which introduces significant complexity. While previous studies have addressed ion transport in PNMs \cite{ahmad_micro-scale_2021} and colloidal suspensions using Lattice Boltzmann methods \cite{qin_lattice_2023}, they did not capture solidification or its impact on drying. Advancing pore-scale models to incorporate sol-gel transitions and solidification remains an interesting and promising direction for future research.

\begin{acknowledgements}
The authors are grateful to Elham Mirzahossein who provided some of the datas from Ref. \cite{le_dizes_castell_visualization_2024}. We also acknowledge Prof. J. Carmeliet and Prof. D. Derome for insighful discussions. 
\end{acknowledgements}

\appendix
\section{Influence of thin films on the drying of a pure liquid} 
\label{sec:Appendix4_TT}

\paragraph*{Building PN without thin films}
The influence of  narrow throats with $w\leq10$ $\mu$m, mimicking the effect of liquid thin films in porous networks is elaborated further here. We  compare the drying kinetic in uniform bimodal PN  to that of uniform \textbf{monomodal} PN with  random distribution of throat widths in the interval $[140,260]$ $\mu$m. In order to ensure that the two PNs have roughly the same volume and surface area at the evaporation side, the uniform porous medium has dimensions of 10 x 40 nodes, leading to $V_{pores} = 4.1$ mm$^3$. Similarly, the replicate PN including a population of very narrow throats (with $w\leq10$ $\mu$m) is compared to its equivalent without them. The size of the  replicate  PN without narrow throats is also 10 $\times$ 40 (and $V_{pores} = 3.5$ mm$^3$) and the other parameters are kept the same. We consider the drying of kinetics of water. 

\begin{figure}[http]
    \centering
    \includegraphics[width=\columnwidth]{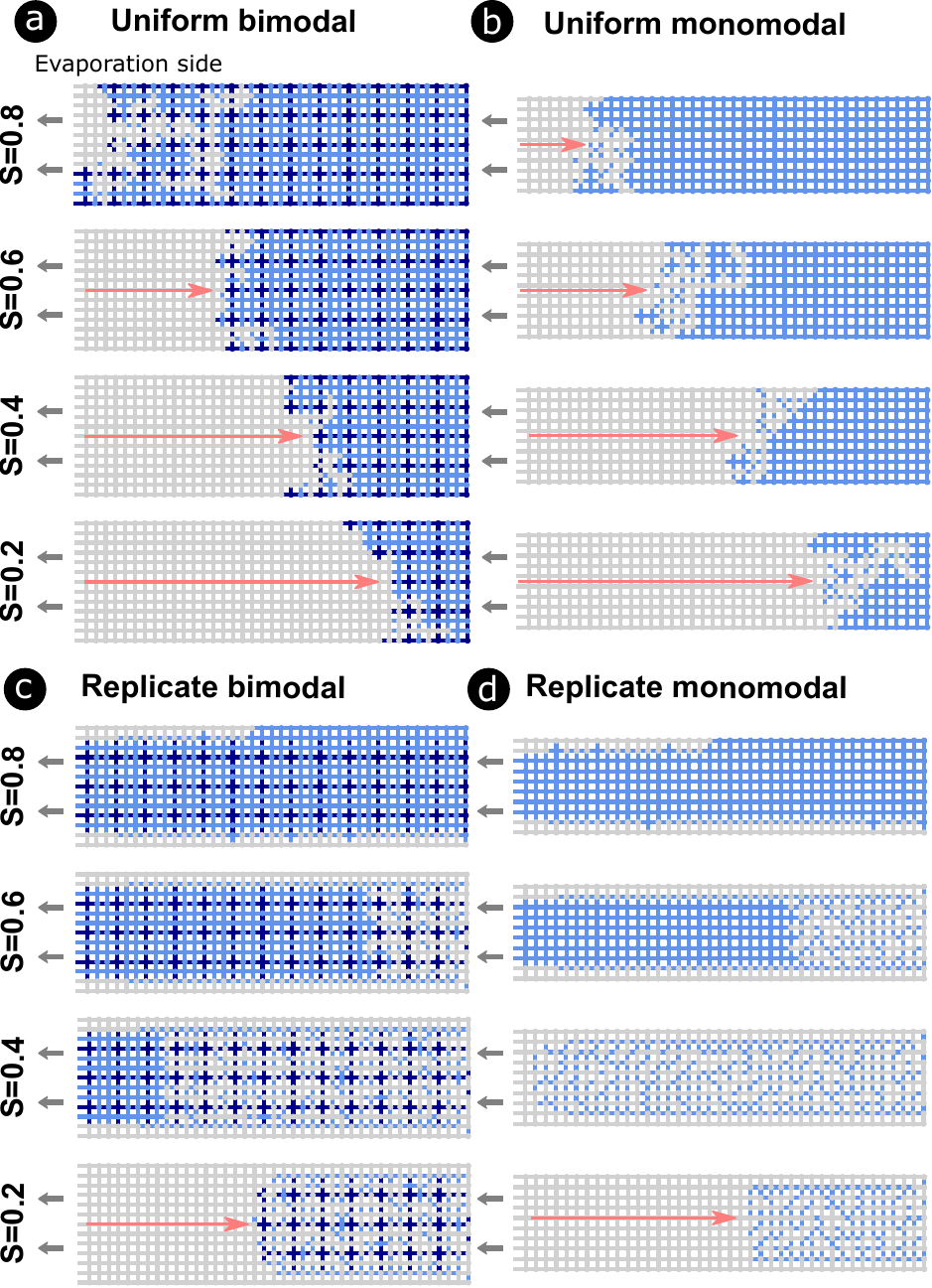}
    \caption{(a) Snapshots at different moments of the drying of water in the uniform PNMs with thin films, compared to the results obtained without thin films in panel (b).  (c) and (d): The same as panels (a) and (b) but for the replicated porous network. The color scheme is similar to the one used in Fig. \ref{Heterogeneities}}
    \label{fig:AppInfluenceTT}
\end{figure}

\begin{figure}[http]
    \centering
    \includegraphics[width=0.9\columnwidth]{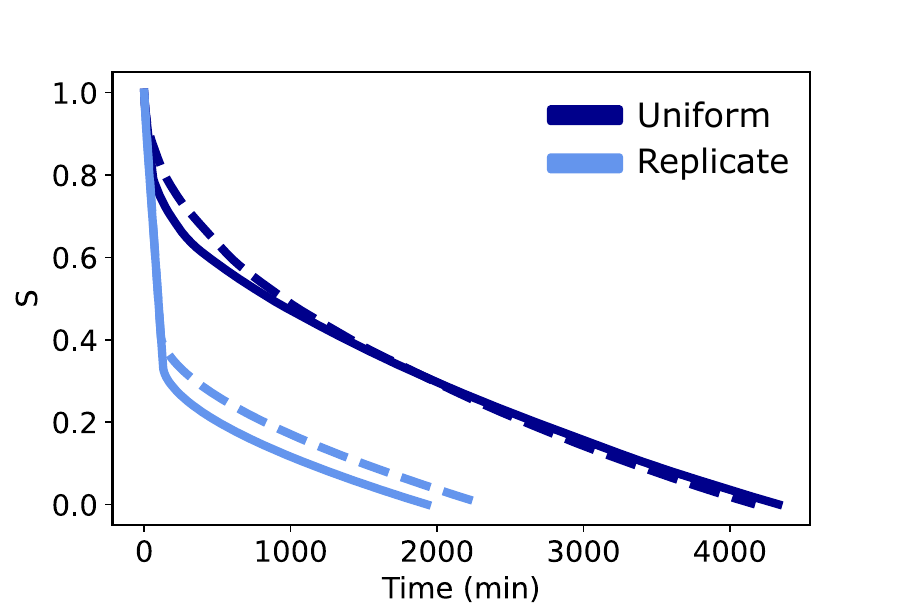}
    \caption{Saturation of water as a function of time in the uniform PNs (dark blue) and in the replicated PN (light blue). The monomodal PNs are the ones corresponding to the dotted lines.}
    \label{fig:AppInfluenceTTDC}
\end{figure}

\paragraph*{Influence on the drying of water}
The air invasion path for the uniform and replicate PNs are depicted in Fig. \ref{fig:AppInfluenceTT}.
For the uniform bimodal and monomodal PNs in Fig. \ref{fig:AppInfluenceTT}(a) and b, no major differences are visible, except that, in the PN without narrow throats with $w\leq10$ $\mu$m, the FRP starts at a slightly higher saturation value. This trend is also observed for the drying curves shown in Fig. \ref{fig:AppInfluenceTTDC}, demonstrating a similar drying kinetics for both PNs. Likewise, in the case of the replicate  PNs shown in Fig. \ref{fig:AppInfluenceTT}(c) and (d), the air invasion path is not significantly influenced by the presence of thin films, as the statistical features of the path remain similar. The FRP also starts at higher saturation value in the case of the   replicate  PN without very narrow throats (see Fig. \ref{fig:AppInfluenceTTDC}). This difference  arises because  most of the liquid remaining in the PN at the end of the CRP is trapped in the narrowest capillary throats, which have low volumes, thus corresponding to low volumes of water while in the  PN, at the end of the CRP, water is trapped randomly in some throats, with bigger mean volumes.

From this discussion, we conclude that the presence of very narrow throat with $w\leq10$ $\mu$m in PNM does not significantly influence the drying kinetics and the air invasion path. Nonetheless, as the thin films have a smaller volume, they trap less liquid that the other throats, leading to FRP that starts at lower saturation values. For the replicated PN, this leads to a better agreement  of drying curves with those of   experiments.


\section{Experimental results for the sol-gel transition in  a single pore} \label{sec:AppRoundCap}
 
\begin{figure*}
    \centering
    \includegraphics[width=0.99\textwidth]{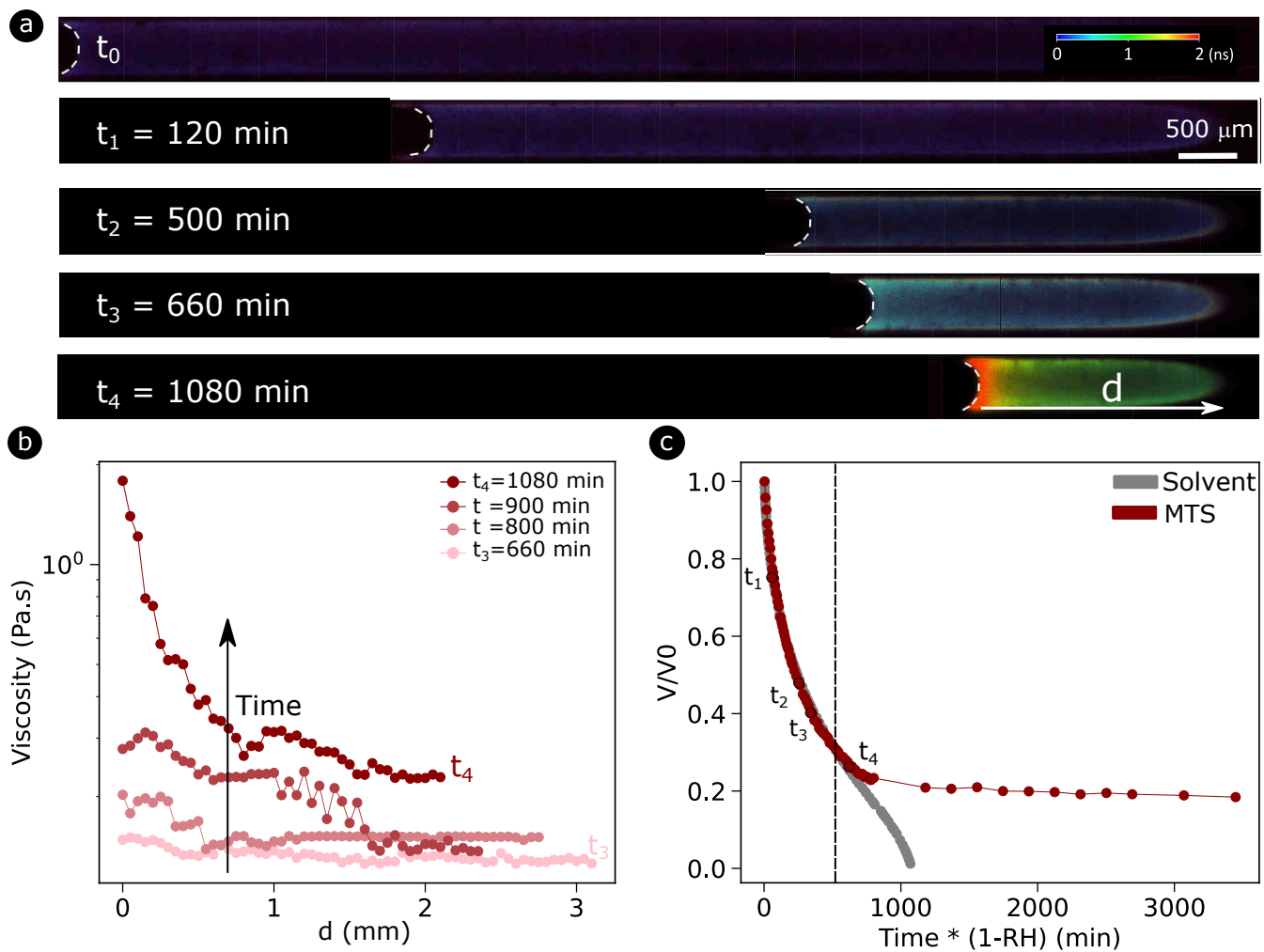}
    \caption{\textbf{Sol-gel transition of MTS in a round capillary open only on the left side, reproduced from Ref. \cite{PhD_Romane}} a) Fluorescence lifetime imaging microscopy images of the gelation process at different evaporation times. The colors correspond to different values of the fluorescence lifetime  (from 0 to 2 ns), as shown in the inset of the figure. b) Calculated viscosity profiles derived from fluorescence lifetime measurements as a function of distance $d$ from the surface of the meniscus, for four different moments between $t_3$ and $t_4$. c) MTS volume change as a function of evaporation time, as calculated from images such as in panel (a) and compared to the solvent. The dashed line corresponds to the moment where the drying kinetics of the MTS deviate from the ones of the solvent}
    \label{fig:roundcap}
\end{figure*}
In Ref. \cite{le_dizes_castell_visualization_2024}, we showed that the  evaporation-induced  sol-gel transition of an organosilicate in a round capillary exhibits two regimes, as reported in Fig. \ref{fig:roundcap}.
In the first period of the drying, the fluid dries at the same rate as the pure solvent, and the viscosity increases homogeneously within the capillary. This periods ends when a layer with an increased viscosity forms at the surface. From that point onward, the evaporation rate is strongly reduced. The transition between the two drying regimes corresponds to the time when the viscosity of the skin at the meniscus reaches a critical value of  $\eta_c \approx 0.22 \pm 0.03$ Pa.s.

As the drying kinetics of gel-forming fluid is similar to that of the solvent when the viscosity  at the meniscus  is below $\eta_c =0.22$ Pa.s, one can consider that the evaporation rate is independent of the viscosity at the meniscus while $\eta <\eta_c$ Pa.s. However, for $\eta > \eta_c$ Pa.s, it is possible to determine a reduced evaporation rate from the experimental data. The evaporation flux $j$ (the amount of mass evaporating per unit of area and time with unit of kg.m$^{-2}$.s$^{-1}$) in the second drying period can be measured experimentally in Fig. \ref{fig:roundcap} as: $j_{ev} = \rho_l \frac{\mathrm{d}\ell}{\mathrm{d}t}$ where $\ell$ is the length between the meniscus and the 'entrance' of the capillary at the evaporation side. $\rho_l$ is the density of the liquid, considered equal to that of water.  For a pure liquid, the vapor diffusive flux is given by Fick's law: $j_d=-D \frac{d \rho_v}{dz}$, where $\rho_v$ is the mass density of vapor and  $D$ is the vapor diffusion coefficient in the air. Using the ideal gas law for the vapor and assuming a quasi-static diffusion, at equilibrium  one  expects a balance between the two fluxes yielding: 
\begin{equation}
   \rho_l \frac{\mathrm{d}\ell}{\mathrm{d}t}= D \frac{M_v}{RT}\frac{\Delta P^{(v)}}{\ell+\delta}
  \label{eq:Stefan}
\end{equation}
where $\Delta P^{(v)}$ is the difference between the vapor pressure at the meniscus and $P_{v\infty}$ the vapor pressure far from the capillary. This equation is generally referred as Stefan's diffusion problem \cite{mitrovic_josef_2012,le_dizes_castell_is_2023} for a pure liquid. 
In the experiments, the fluid is not a pure liquid and the  evaporating solvent is a mixture of ethanol and water. Nonetheless, the skin formation ({\it i.e.}, $\eta > \eta_c$  ) occurs for a saturation $S \approx 0.25$, meaning than most of the liquid has already evaporated. It is reasonable to approximate that, for this saturation, the solvent is composed only of water, as the ethanol is more volatile and evaporates faster. One can then estimate, from Eq. \eqref{eq:Stefan} for water and experimental results, $\Delta P^{(v)}$ after the skin formation. From Fig. \ref{fig:roundcap}, with $\frac{\mathrm{d}\ell}{\mathrm{d}t} = 10^{-6}$ kg.m$^{-2}$.s$^{-1}$ and $\ell=7$ mm, we can determine $\Delta P^{(v)} = 40$ Pa.  For $\eta>\eta_c$   at the meniscus, the drying rate in the round capillary can thus be approximated by considering that the vapor pressure above the meniscus is equal to $P^{(v)} = \Delta P^{(v)} + P^{(v)}_{\infty} \approx 1100$ Pa as $P^{(v)}_{\infty}=RH*P^{(v)}_{sat}= 1053$. \\
\bibliography{PNMref}

\begin{thebibliography}{39}%
\makeatletter
\providecommand \@ifxundefined [1]{%
 \@ifx{#1\undefined}
}%
\providecommand \@ifnum [1]{%
 \ifnum #1\expandafter \@firstoftwo
 \else \expandafter \@secondoftwo
 \fi
}%
\providecommand \@ifx [1]{%
 \ifx #1\expandafter \@firstoftwo
 \else \expandafter \@secondoftwo
 \fi
}%
\providecommand \natexlab [1]{#1}%
\providecommand \enquote  [1]{``#1''}%
\providecommand \bibnamefont  [1]{#1}%
\providecommand \bibfnamefont [1]{#1}%
\providecommand \citenamefont [1]{#1}%
\providecommand \href@noop [0]{\@secondoftwo}%
\providecommand \href [0]{\begingroup \@sanitize@url \@href}%
\providecommand \@href[1]{\@@startlink{#1}\@@href}%
\providecommand \@@href[1]{\endgroup#1\@@endlink}%
\providecommand \@sanitize@url [0]{\catcode `\\12\catcode `\$12\catcode `\&12\catcode `\#12\catcode `\^12\catcode `\_12\catcode `\%12\relax}%
\providecommand \@@startlink[1]{}%
\providecommand \@@endlink[0]{}%
\providecommand \url  [0]{\begingroup\@sanitize@url \@url }%
\providecommand \@url [1]{\endgroup\@href {#1}{\urlprefix }}%
\providecommand \urlprefix  [0]{URL }%
\providecommand \Eprint [0]{\href }%
\providecommand \doibase [0]{http://dx.doi.org/}%
\providecommand \selectlanguage [0]{\@gobble}%
\providecommand \bibinfo  [0]{\@secondoftwo}%
\providecommand \bibfield  [0]{\@secondoftwo}%
\providecommand \translation [1]{[#1]}%
\providecommand \BibitemOpen [0]{}%
\providecommand \bibitemStop [0]{}%
\providecommand \bibitemNoStop [0]{.\EOS\space}%
\providecommand \EOS [0]{\spacefactor3000\relax}%
\providecommand \BibitemShut  [1]{\csname bibitem#1\endcsname}%
\let\auto@bib@innerbib\@empty
\bibitem [{\citenamefont {Le~Dizès~Castell}\ \emph {et~al.}(2024{\natexlab{a}})\citenamefont {Le~Dizès~Castell}, \citenamefont {Mirzahossein}, \citenamefont {Grzelka}, \citenamefont {Jabbari-Farouji}, \citenamefont {Bonn},\ and\ \citenamefont {Shahidzadeh}}]{le_dizes_castell_visualization_2024}%
  \BibitemOpen
  \bibfield  {author} {\bibinfo {author} {\bibfnamefont {R.}~\bibnamefont {Le~Dizès~Castell}}, \bibinfo {author} {\bibfnamefont {E.}~\bibnamefont {Mirzahossein}}, \bibinfo {author} {\bibfnamefont {M.}~\bibnamefont {Grzelka}}, \bibinfo {author} {\bibfnamefont {S.}~\bibnamefont {Jabbari-Farouji}}, \bibinfo {author} {\bibfnamefont {D.}~\bibnamefont {Bonn}}, \ and\ \bibinfo {author} {\bibfnamefont {N.}~\bibnamefont {Shahidzadeh}},\ }\href {\doibase 10.1021/acs.jpclett.3c02634} {\bibfield  {journal} {\bibinfo  {journal} {The Journal of Physical Chemistry Letters}\ }\textbf {\bibinfo {volume} {15}},\ \bibinfo {pages} {628} (\bibinfo {year} {2024}{\natexlab{a}})},\ \bibinfo {note} {publisher: American Chemical Society}\BibitemShut {NoStop}%
\bibitem [{\citenamefont {Okuzono}\ \emph {et~al.}(2006)\citenamefont {Okuzono}, \citenamefont {Ozawa},\ and\ \citenamefont {Doi}}]{okuzono_simple_2006}%
  \BibitemOpen
  \bibfield  {author} {\bibinfo {author} {\bibfnamefont {T.}~\bibnamefont {Okuzono}}, \bibinfo {author} {\bibfnamefont {K.}~\bibnamefont {Ozawa}}, \ and\ \bibinfo {author} {\bibfnamefont {M.}~\bibnamefont {Doi}},\ }\href {\doibase 10.1103/PhysRevLett.97.136103} {\bibfield  {journal} {\bibinfo  {journal} {Physical Review Letters}\ }\textbf {\bibinfo {volume} {97}},\ \bibinfo {pages} {136103} (\bibinfo {year} {2006})}\BibitemShut {NoStop}%
\bibitem [{\citenamefont {Talini}\ and\ \citenamefont {Lequeux}(2023)}]{talini_formation_2023}%
  \BibitemOpen
  \bibfield  {author} {\bibinfo {author} {\bibfnamefont {L.}~\bibnamefont {Talini}}\ and\ \bibinfo {author} {\bibfnamefont {F.}~\bibnamefont {Lequeux}},\ }\href {\doibase 10.1039/D3SM00522D} {\bibfield  {journal} {\bibinfo  {journal} {Soft Matter}\ }\textbf {\bibinfo {volume} {19}},\ \bibinfo {pages} {5835} (\bibinfo {year} {2023})}\BibitemShut {NoStop}%
\bibitem [{\citenamefont {Salmon}\ \emph {et~al.}(2017)\citenamefont {Salmon}, \citenamefont {Doumenc},\ and\ \citenamefont {Guerrier}}]{salmon_humidity-insensitive_2017}%
  \BibitemOpen
  \bibfield  {author} {\bibinfo {author} {\bibfnamefont {J.-B.}\ \bibnamefont {Salmon}}, \bibinfo {author} {\bibfnamefont {F.}~\bibnamefont {Doumenc}}, \ and\ \bibinfo {author} {\bibfnamefont {B.}~\bibnamefont {Guerrier}},\ }\href {\doibase 10.1103/PhysRevE.96.032612} {\bibfield  {journal} {\bibinfo  {journal} {Physical Review E}\ }\textbf {\bibinfo {volume} {96}},\ \bibinfo {pages} {032612} (\bibinfo {year} {2017})}\BibitemShut {NoStop}%
\bibitem [{\citenamefont {Faiyas}\ \emph {et~al.}(2017)\citenamefont {Faiyas}, \citenamefont {Erich}, \citenamefont {Huinink},\ and\ \citenamefont {Adan}}]{faiyas_transport_2017}%
  \BibitemOpen
  \bibfield  {author} {\bibinfo {author} {\bibfnamefont {A.~P.~A.}\ \bibnamefont {Faiyas}}, \bibinfo {author} {\bibfnamefont {S.~J.~F.}\ \bibnamefont {Erich}}, \bibinfo {author} {\bibfnamefont {H.~P.}\ \bibnamefont {Huinink}}, \ and\ \bibinfo {author} {\bibfnamefont {O.~C.~G.}\ \bibnamefont {Adan}},\ }\href {\doibase 10.1080/07373937.2017.1283515} {\bibfield  {journal} {\bibinfo  {journal} {Drying Technology}\ }\textbf {\bibinfo {volume} {35}},\ \bibinfo {pages} {1874} (\bibinfo {year} {2017})}\BibitemShut {NoStop}%
\bibitem [{\citenamefont {Brinker}\ \emph {et~al.}(1992)\citenamefont {Brinker}, \citenamefont {Hurd}, \citenamefont {Schunk}, \citenamefont {Frye},\ and\ \citenamefont {Ashley}}]{brinker_review_1992}%
  \BibitemOpen
  \bibfield  {author} {\bibinfo {author} {\bibfnamefont {C.~J.}\ \bibnamefont {Brinker}}, \bibinfo {author} {\bibfnamefont {A.~J.}\ \bibnamefont {Hurd}}, \bibinfo {author} {\bibfnamefont {P.~R.}\ \bibnamefont {Schunk}}, \bibinfo {author} {\bibfnamefont {G.~C.}\ \bibnamefont {Frye}}, \ and\ \bibinfo {author} {\bibfnamefont {C.~S.}\ \bibnamefont {Ashley}},\ }\href {\doibase 10.1016/S0022-3093(05)80653-2} {\bibfield  {journal} {\bibinfo  {journal} {Journal of Non-Crystalline Solids}\ }\bibinfo {series} {Advanced {Materials} from {Gels}},\ \textbf {\bibinfo {volume} {147-148}},\ \bibinfo {pages} {424} (\bibinfo {year} {1992})}\BibitemShut {NoStop}%
\bibitem [{\citenamefont {Pauchard}\ and\ \citenamefont {Allain}(2003{\natexlab{a}})}]{pauchard_stable_2003}%
  \BibitemOpen
  \bibfield  {author} {\bibinfo {author} {\bibfnamefont {L.}~\bibnamefont {Pauchard}}\ and\ \bibinfo {author} {\bibfnamefont {C.}~\bibnamefont {Allain}},\ }\href {\doibase 10.1103/PhysRevE.68.052801} {\bibfield  {journal} {\bibinfo  {journal} {Physical Review E}\ }\textbf {\bibinfo {volume} {68}},\ \bibinfo {pages} {052801} (\bibinfo {year} {2003}{\natexlab{a}})}\BibitemShut {NoStop}%
\bibitem [{\citenamefont {Pauchard}\ and\ \citenamefont {Allain}(2003{\natexlab{b}})}]{pauchard_buckling_2003}%
  \BibitemOpen
  \bibfield  {author} {\bibinfo {author} {\bibfnamefont {L.}~\bibnamefont {Pauchard}}\ and\ \bibinfo {author} {\bibfnamefont {C.}~\bibnamefont {Allain}},\ }\href {\doibase 10.1209/epl/i2003-00457-7} {\bibfield  {journal} {\bibinfo  {journal} {Europhysics Letters}\ }\textbf {\bibinfo {volume} {62}},\ \bibinfo {pages} {897} (\bibinfo {year} {2003}{\natexlab{b}})}\BibitemShut {NoStop}%
\bibitem [{\citenamefont {Huisman}\ \emph {et~al.}(2023)\citenamefont {Huisman}, \citenamefont {Digard}, \citenamefont {Poon},\ and\ \citenamefont {Titmuss}}]{huisman_evaporation_2023}%
  \BibitemOpen
  \bibfield  {author} {\bibinfo {author} {\bibfnamefont {M.}~\bibnamefont {Huisman}}, \bibinfo {author} {\bibfnamefont {P.}~\bibnamefont {Digard}}, \bibinfo {author} {\bibfnamefont {W.~C.}\ \bibnamefont {Poon}}, \ and\ \bibinfo {author} {\bibfnamefont {S.}~\bibnamefont {Titmuss}},\ }\href {\doibase 10.1103/PhysRevLett.131.248102} {\bibfield  {journal} {\bibinfo  {journal} {Physical Review Letters}\ }\textbf {\bibinfo {volume} {131}},\ \bibinfo {pages} {248102} (\bibinfo {year} {2023})}\BibitemShut {NoStop}%
\bibitem [{\citenamefont {Cairncross}\ \emph {et~al.}(1996)\citenamefont {Cairncross}, \citenamefont {Francis},\ and\ \citenamefont {Scriven}}]{cairncross_predicting_1996}%
  \BibitemOpen
  \bibfield  {author} {\bibinfo {author} {\bibfnamefont {R.~A.}\ \bibnamefont {Cairncross}}, \bibinfo {author} {\bibfnamefont {L.~F.}\ \bibnamefont {Francis}}, \ and\ \bibinfo {author} {\bibfnamefont {L.~E.}\ \bibnamefont {Scriven}},\ }\href {\doibase 10.1002/aic.690420107} {\bibfield  {journal} {\bibinfo  {journal} {AIChE Journal}\ }\textbf {\bibinfo {volume} {42}},\ \bibinfo {pages} {55} (\bibinfo {year} {1996})}\BibitemShut {NoStop}%
\bibitem [{\citenamefont {Bühler}\ \emph {et~al.}(2013)\citenamefont {Bühler}, \citenamefont {Zurbriggen}, \citenamefont {Pieles}, \citenamefont {Huwiler},\ and\ \citenamefont {Raso}}]{buhler_dynamics_2013}%
  \BibitemOpen
  \bibfield  {author} {\bibinfo {author} {\bibfnamefont {T.}~\bibnamefont {Bühler}}, \bibinfo {author} {\bibfnamefont {R.}~\bibnamefont {Zurbriggen}}, \bibinfo {author} {\bibfnamefont {U.}~\bibnamefont {Pieles}}, \bibinfo {author} {\bibfnamefont {L.}~\bibnamefont {Huwiler}}, \ and\ \bibinfo {author} {\bibfnamefont {R.~A.}\ \bibnamefont {Raso}},\ }\href {\doibase 10.1016/j.cemconcomp.2012.10.008} {\bibfield  {journal} {\bibinfo  {journal} {Cement and Concrete Composites}\ }\textbf {\bibinfo {volume} {37}},\ \bibinfo {pages} {161} (\bibinfo {year} {2013})}\BibitemShut {NoStop}%
\bibitem [{\citenamefont {Faiyas}\ \emph {et~al.}(2015)\citenamefont {Faiyas}, \citenamefont {Erich}, \citenamefont {van Soestbergen}, \citenamefont {Huinink}, \citenamefont {Adan},\ and\ \citenamefont {Nijland}}]{faiyas_how_2015}%
  \BibitemOpen
  \bibfield  {author} {\bibinfo {author} {\bibfnamefont {A.~P.~A.}\ \bibnamefont {Faiyas}}, \bibinfo {author} {\bibfnamefont {S.~J.~F.}\ \bibnamefont {Erich}}, \bibinfo {author} {\bibfnamefont {M.}~\bibnamefont {van Soestbergen}}, \bibinfo {author} {\bibfnamefont {H.~P.}\ \bibnamefont {Huinink}}, \bibinfo {author} {\bibfnamefont {O.~C.~G.}\ \bibnamefont {Adan}}, \ and\ \bibinfo {author} {\bibfnamefont {T.~G.}\ \bibnamefont {Nijland}},\ }\href {\doibase 10.1016/j.ces.2014.11.054} {\bibfield  {journal} {\bibinfo  {journal} {Chemical Engineering Science}\ }\textbf {\bibinfo {volume} {123}},\ \bibinfo {pages} {620} (\bibinfo {year} {2015})}\BibitemShut {NoStop}%
\bibitem [{\citenamefont {Or}\ \emph {et~al.}(2007)\citenamefont {Or}, \citenamefont {Phutane},\ and\ \citenamefont {Dechesne}}]{or_extracellular_2007}%
  \BibitemOpen
  \bibfield  {author} {\bibinfo {author} {\bibfnamefont {D.}~\bibnamefont {Or}}, \bibinfo {author} {\bibfnamefont {S.}~\bibnamefont {Phutane}}, \ and\ \bibinfo {author} {\bibfnamefont {A.}~\bibnamefont {Dechesne}},\ }\href {\doibase 10.2136/vzj2006.0080} {\bibfield  {journal} {\bibinfo  {journal} {Vadose Zone Journal}\ }\textbf {\bibinfo {volume} {6}},\ \bibinfo {pages} {298} (\bibinfo {year} {2007})}\BibitemShut {NoStop}%
\bibitem [{\citenamefont {Deng}\ \emph {et~al.}(2015)\citenamefont {Deng}, \citenamefont {Orner}, \citenamefont {Chau}, \citenamefont {Anderson}, \citenamefont {Kadilak}, \citenamefont {Rubinstein}, \citenamefont {Bouchillon}, \citenamefont {Goodwin}, \citenamefont {Gage},\ and\ \citenamefont {Shor}}]{deng_synergistic_2015}%
  \BibitemOpen
  \bibfield  {author} {\bibinfo {author} {\bibfnamefont {J.}~\bibnamefont {Deng}}, \bibinfo {author} {\bibfnamefont {E.~P.}\ \bibnamefont {Orner}}, \bibinfo {author} {\bibfnamefont {J.~F.}\ \bibnamefont {Chau}}, \bibinfo {author} {\bibfnamefont {E.~M.}\ \bibnamefont {Anderson}}, \bibinfo {author} {\bibfnamefont {A.~L.}\ \bibnamefont {Kadilak}}, \bibinfo {author} {\bibfnamefont {R.~L.}\ \bibnamefont {Rubinstein}}, \bibinfo {author} {\bibfnamefont {G.~M.}\ \bibnamefont {Bouchillon}}, \bibinfo {author} {\bibfnamefont {R.~A.}\ \bibnamefont {Goodwin}}, \bibinfo {author} {\bibfnamefont {D.~J.}\ \bibnamefont {Gage}}, \ and\ \bibinfo {author} {\bibfnamefont {L.~M.}\ \bibnamefont {Shor}},\ }\href {\doibase 10.1016/j.soilbio.2014.12.006} {\bibfield  {journal} {\bibinfo  {journal} {Soil Biology and Biochemistry}\ }\textbf {\bibinfo {volume} {83}},\ \bibinfo {pages} {116} (\bibinfo {year} {2015})}\BibitemShut {NoStop}%
\bibitem [{\citenamefont {Guo}\ \emph {et~al.}(2018)\citenamefont {Guo}, \citenamefont {Furrer}, \citenamefont {Kadilak}, \citenamefont {Hinestroza}, \citenamefont {Gage}, \citenamefont {Cho},\ and\ \citenamefont {Shor}}]{guo_bacterial_2018}%
  \BibitemOpen
  \bibfield  {author} {\bibinfo {author} {\bibfnamefont {Y.-S.}\ \bibnamefont {Guo}}, \bibinfo {author} {\bibfnamefont {J.~M.}\ \bibnamefont {Furrer}}, \bibinfo {author} {\bibfnamefont {A.~L.}\ \bibnamefont {Kadilak}}, \bibinfo {author} {\bibfnamefont {H.~F.}\ \bibnamefont {Hinestroza}}, \bibinfo {author} {\bibfnamefont {D.~J.}\ \bibnamefont {Gage}}, \bibinfo {author} {\bibfnamefont {Y.~K.}\ \bibnamefont {Cho}}, \ and\ \bibinfo {author} {\bibfnamefont {L.~M.}\ \bibnamefont {Shor}},\ }\href {\doibase 10.3389/fenvs.2018.00093} {\bibfield  {journal} {\bibinfo  {journal} {Frontiers in Environmental Science}\ }\textbf {\bibinfo {volume} {6}} (\bibinfo {year} {2018}),\ 10.3389/fenvs.2018.00093}\BibitemShut {NoStop}%
\bibitem [{\citenamefont {Wheeler}(2005)}]{wheeler_alkoxysilanes_2005}%
  \BibitemOpen
  \bibfield  {author} {\bibinfo {author} {\bibfnamefont {G.}~\bibnamefont {Wheeler}},\ }\href@noop {} {\emph {\bibinfo {title} {Alkoxysilanes and the {Consolidation} of {Stone}}}}\ (\bibinfo  {publisher} {Getty Publications},\ \bibinfo {year} {2005})\BibitemShut {NoStop}%
\bibitem [{\citenamefont {Le~Dizès~Castell}\ \emph {et~al.}(2024{\natexlab{b}})\citenamefont {Le~Dizès~Castell}, \citenamefont {Pel}, \citenamefont {Chekai}, \citenamefont {Derluyn}, \citenamefont {Scheel}, \citenamefont {Jabbari-Farouji},\ and\ \citenamefont {Shahidzadeh}}]{le_dizes_castell_sol-gel_2024}%
  \BibitemOpen
  \bibfield  {author} {\bibinfo {author} {\bibfnamefont {R.}~\bibnamefont {Le~Dizès~Castell}}, \bibinfo {author} {\bibfnamefont {L.}~\bibnamefont {Pel}}, \bibinfo {author} {\bibfnamefont {T.}~\bibnamefont {Chekai}}, \bibinfo {author} {\bibfnamefont {H.}~\bibnamefont {Derluyn}}, \bibinfo {author} {\bibfnamefont {M.}~\bibnamefont {Scheel}}, \bibinfo {author} {\bibfnamefont {S.}~\bibnamefont {Jabbari-Farouji}}, \ and\ \bibinfo {author} {\bibfnamefont {N.}~\bibnamefont {Shahidzadeh}},\ }\href {\doibase 10.1103/PhysRevApplied.21.034049} {\bibfield  {journal} {\bibinfo  {journal} {Physical Review Applied}\ }\textbf {\bibinfo {volume} {21}},\ \bibinfo {pages} {034049} (\bibinfo {year} {2024}{\natexlab{b}})}\BibitemShut {NoStop}%
\bibitem [{\citenamefont {Wilkinson}\ and\ \citenamefont {Willemsen}(1983)}]{wilkinson_invasion_1983}%
  \BibitemOpen
  \bibfield  {author} {\bibinfo {author} {\bibfnamefont {D.}~\bibnamefont {Wilkinson}}\ and\ \bibinfo {author} {\bibfnamefont {J.~F.}\ \bibnamefont {Willemsen}},\ }\href {\doibase 10.1088/0305-4470/16/14/028} {\bibfield  {journal} {\bibinfo  {journal} {Journal of Physics A: Mathematical and General}\ }\textbf {\bibinfo {volume} {16}},\ \bibinfo {pages} {3365} (\bibinfo {year} {1983})}\BibitemShut {NoStop}%
\bibitem [{\citenamefont {Wilkinson}(1986)}]{wilkinson_percolation_1986}%
  \BibitemOpen
  \bibfield  {author} {\bibinfo {author} {\bibfnamefont {D.}~\bibnamefont {Wilkinson}},\ }\href {\doibase 10.1103/PhysRevA.34.1380} {\bibfield  {journal} {\bibinfo  {journal} {Physical Review A}\ }\textbf {\bibinfo {volume} {34}},\ \bibinfo {pages} {1380} (\bibinfo {year} {1986})}\BibitemShut {NoStop}%
\bibitem [{\citenamefont {Nowicki}\ \emph {et~al.}(1992)\citenamefont {Nowicki}, \citenamefont {Davis},\ and\ \citenamefont {Scriven}}]{nowicki_microscopic_1992}%
  \BibitemOpen
  \bibfield  {author} {\bibinfo {author} {\bibfnamefont {S.~C.}\ \bibnamefont {Nowicki}}, \bibinfo {author} {\bibfnamefont {H.~T.}\ \bibnamefont {Davis}}, \ and\ \bibinfo {author} {\bibfnamefont {L.~E.}\ \bibnamefont {Scriven}},\ }\href {\doibase 10.1080/07373939208916488} {\bibfield  {journal} {\bibinfo  {journal} {Drying Technology}\ }\textbf {\bibinfo {volume} {10}},\ \bibinfo {pages} {925} (\bibinfo {year} {1992})}\BibitemShut {NoStop}%
\bibitem [{\citenamefont {Prat}(1993)}]{prat_percolation_1993}%
  \BibitemOpen
  \bibfield  {author} {\bibinfo {author} {\bibfnamefont {M.}~\bibnamefont {Prat}},\ }\href {\doibase 10.1016/0301-9322(93)90096-D} {\bibfield  {journal} {\bibinfo  {journal} {International Journal of Multiphase Flow}\ }\textbf {\bibinfo {volume} {19}},\ \bibinfo {pages} {691} (\bibinfo {year} {1993})}\BibitemShut {NoStop}%
\bibitem [{\citenamefont {Yiotis}\ \emph {et~al.}(2001)\citenamefont {Yiotis}, \citenamefont {Stubos}, \citenamefont {Boudouvis},\ and\ \citenamefont {Yortsos}}]{yiotis_2-d_2001}%
  \BibitemOpen
  \bibfield  {author} {\bibinfo {author} {\bibfnamefont {A.~G.}\ \bibnamefont {Yiotis}}, \bibinfo {author} {\bibfnamefont {A.~K.}\ \bibnamefont {Stubos}}, \bibinfo {author} {\bibfnamefont {A.~G.}\ \bibnamefont {Boudouvis}}, \ and\ \bibinfo {author} {\bibfnamefont {Y.~C.}\ \bibnamefont {Yortsos}},\ }\href@noop {} {\bibfield  {journal} {\bibinfo  {journal} {Advances in Water Resources}\ }\textbf {\bibinfo {volume} {24}},\ \bibinfo {pages} {437} (\bibinfo {year} {2001})}\BibitemShut {NoStop}%
\bibitem [{\citenamefont {Metzger}\ \emph {et~al.}(2007)\citenamefont {Metzger}, \citenamefont {Irawan},\ and\ \citenamefont {Tsotsas}}]{metzger_isothermal_2007}%
  \BibitemOpen
  \bibfield  {author} {\bibinfo {author} {\bibfnamefont {T.}~\bibnamefont {Metzger}}, \bibinfo {author} {\bibfnamefont {A.}~\bibnamefont {Irawan}}, \ and\ \bibinfo {author} {\bibfnamefont {E.}~\bibnamefont {Tsotsas}},\ }\href {\doibase 10.1080/07373930601152640} {\bibfield  {journal} {\bibinfo  {journal} {Drying Technology}\ }\textbf {\bibinfo {volume} {25}},\ \bibinfo {pages} {49} (\bibinfo {year} {2007})}\BibitemShut {NoStop}%
\bibitem [{\citenamefont {Prat}(1995)}]{prat_isothermal_1995}%
  \BibitemOpen
  \bibfield  {author} {\bibinfo {author} {\bibfnamefont {M.}~\bibnamefont {Prat}},\ }\href {\doibase 10.1016/0301-9322(95)00022-P} {\bibfield  {journal} {\bibinfo  {journal} {International Journal of Multiphase Flow}\ }\textbf {\bibinfo {volume} {21}},\ \bibinfo {pages} {875} (\bibinfo {year} {1995})}\BibitemShut {NoStop}%
\bibitem [{\citenamefont {Plourde}\ and\ \citenamefont {Prat}(2003)}]{plourde_pore_2003}%
  \BibitemOpen
  \bibfield  {author} {\bibinfo {author} {\bibfnamefont {F.}~\bibnamefont {Plourde}}\ and\ \bibinfo {author} {\bibfnamefont {M.}~\bibnamefont {Prat}},\ }\href {\doibase 10.1016/S0017-9310(02)00391-5} {\bibfield  {journal} {\bibinfo  {journal} {International Journal of Heat and Mass Transfer}\ }\textbf {\bibinfo {volume} {46}},\ \bibinfo {pages} {1293} (\bibinfo {year} {2003})}\BibitemShut {NoStop}%
\bibitem [{\citenamefont {Hoshen}\ and\ \citenamefont {Kopelman}(1976)}]{hoshen_percolation_1976}%
  \BibitemOpen
  \bibfield  {author} {\bibinfo {author} {\bibfnamefont {J.}~\bibnamefont {Hoshen}}\ and\ \bibinfo {author} {\bibfnamefont {R.}~\bibnamefont {Kopelman}},\ }\href {\doibase 10.1103/PhysRevB.14.3438} {\bibfield  {journal} {\bibinfo  {journal} {Physical Review B}\ }\textbf {\bibinfo {volume} {14}},\ \bibinfo {pages} {3438} (\bibinfo {year} {1976})}\BibitemShut {NoStop}%
\bibitem [{\citenamefont {Versteeg}\ and\ \citenamefont {Malalasekera}(2007)}]{versteeg_introduction_2007}%
  \BibitemOpen
  \bibfield  {author} {\bibinfo {author} {\bibfnamefont {H.~K.}\ \bibnamefont {Versteeg}}\ and\ \bibinfo {author} {\bibfnamefont {W.}~\bibnamefont {Malalasekera}},\ }\href@noop {} {\emph {\bibinfo {title} {An introduction to computational fluid dynamics: the finite volume method}}},\ \bibinfo {edition} {2nd}\ ed.\ (\bibinfo  {publisher} {Pearson Education Ltd},\ \bibinfo {address} {Harlow, England ; New York},\ \bibinfo {year} {2007})\BibitemShut {NoStop}%
\bibitem [{\citenamefont {Le~Dizes~Castell}(2024)}]{PhD_Romane}%
  \BibitemOpen
  \bibfield  {author} {\bibinfo {author} {\bibfnamefont {R.}~\bibnamefont {Le~Dizes~Castell}},\ }\emph {\bibinfo {title} {Sol-gel transition in porous media: {Interplay} of drying and wetting}},\ \href {https://dare.uva.nl/search?field1=keyword;value1=le%20dizes;docsPerPage=1;startDoc=1} {Ph.D. thesis},\ \bibinfo  {school} {Universiteit van Amsterdam} (\bibinfo {year} {2024})\BibitemShut {NoStop}%
\bibitem [{\citenamefont {Yiotis}\ \emph {et~al.}(2003)\citenamefont {Yiotis}, \citenamefont {Boudouvis}, \citenamefont {Stubos}, \citenamefont {Tsimpanogiannis},\ and\ \citenamefont {Yortsos}}]{yiotis_effect_2003}%
  \BibitemOpen
  \bibfield  {author} {\bibinfo {author} {\bibfnamefont {A.~G.}\ \bibnamefont {Yiotis}}, \bibinfo {author} {\bibfnamefont {A.~G.}\ \bibnamefont {Boudouvis}}, \bibinfo {author} {\bibfnamefont {A.~K.}\ \bibnamefont {Stubos}}, \bibinfo {author} {\bibfnamefont {I.~N.}\ \bibnamefont {Tsimpanogiannis}}, \ and\ \bibinfo {author} {\bibfnamefont {Y.~C.}\ \bibnamefont {Yortsos}},\ }\href {\doibase 10.1103/PhysRevE.68.037303} {\bibfield  {journal} {\bibinfo  {journal} {Physical Review E}\ }\textbf {\bibinfo {volume} {68}},\ \bibinfo {pages} {037303} (\bibinfo {year} {2003})}\BibitemShut {NoStop}%
\bibitem [{\citenamefont {Laurindo}\ and\ \citenamefont {Prat}(1998)}]{laurindo_numerical_1998}%
  \BibitemOpen
  \bibfield  {author} {\bibinfo {author} {\bibfnamefont {J.~B.}\ \bibnamefont {Laurindo}}\ and\ \bibinfo {author} {\bibfnamefont {M.}~\bibnamefont {Prat}},\ }\href {\doibase 10.1016/S0009-2509(97)00348-5} {\bibfield  {journal} {\bibinfo  {journal} {Chemical Engineering Science}\ }\textbf {\bibinfo {volume} {53}},\ \bibinfo {pages} {2257} (\bibinfo {year} {1998})}\BibitemShut {NoStop}%
\bibitem [{\citenamefont {Coussot}(2000)}]{coussot_scaling_2000}%
  \BibitemOpen
  \bibfield  {author} {\bibinfo {author} {\bibfnamefont {P.}~\bibnamefont {Coussot}},\ }\href {\doibase 10.1007/s100510051160} {\bibfield  {journal} {\bibinfo  {journal} {The European Physical Journal B}\ }\textbf {\bibinfo {volume} {15}},\ \bibinfo {pages} {557} (\bibinfo {year} {2000})}\BibitemShut {NoStop}%
\bibitem [{\citenamefont {Prat}(2002)}]{prat_recent_2002}%
  \BibitemOpen
  \bibfield  {author} {\bibinfo {author} {\bibfnamefont {M.}~\bibnamefont {Prat}},\ }\href {\doibase 10.1016/S1385-8947(01)00283-2} {\bibfield  {journal} {\bibinfo  {journal} {Chemical Engineering Journal}\ }\textbf {\bibinfo {volume} {86}},\ \bibinfo {pages} {153} (\bibinfo {year} {2002})}\BibitemShut {NoStop}%
\bibitem [{\citenamefont {Freitas}\ and\ \citenamefont {Prat}(2000)}]{freitas_pore_2000}%
  \BibitemOpen
  \bibfield  {author} {\bibinfo {author} {\bibfnamefont {D.~S.}\ \bibnamefont {Freitas}}\ and\ \bibinfo {author} {\bibfnamefont {M.}~\bibnamefont {Prat}},\ }\href {\doibase 10.1023/A:1006651524722} {\bibfield  {journal} {\bibinfo  {journal} {Transport in Porous Media}\ }\textbf {\bibinfo {volume} {40}},\ \bibinfo {pages} {1} (\bibinfo {year} {2000})}\BibitemShut {NoStop}%
\bibitem [{\citenamefont {Brinker}\ and\ \citenamefont {Scherer}(2013)}]{brinker_sol-gel_2013}%
  \BibitemOpen
  \bibfield  {author} {\bibinfo {author} {\bibfnamefont {C.~J.}\ \bibnamefont {Brinker}}\ and\ \bibinfo {author} {\bibfnamefont {G.~W.}\ \bibnamefont {Scherer}},\ }\href@noop {} {\emph {\bibinfo {title} {Sol-Gel Science: The Physics and Chemistry of Sol-Gel Processing}}}\ (\bibinfo  {publisher} {Academic Press},\ \bibinfo {year} {2013})\ \bibinfo {note} {google-Books-ID: CND1BAAAQBAJ}\BibitemShut {NoStop}%
\bibitem [{\citenamefont {Fei}\ \emph {et~al.}(2024)\citenamefont {Fei}, \citenamefont {Derome},\ and\ \citenamefont {Carmeliet}}]{fei_pore-scale_2024}%
  \BibitemOpen
  \bibfield  {author} {\bibinfo {author} {\bibfnamefont {L.}~\bibnamefont {Fei}}, \bibinfo {author} {\bibfnamefont {D.}~\bibnamefont {Derome}}, \ and\ \bibinfo {author} {\bibfnamefont {J.}~\bibnamefont {Carmeliet}},\ }\href {\doibase 10.1017/jfm.2024.138} {\bibfield  {journal} {\bibinfo  {journal} {Journal of Fluid Mechanics}\ }\textbf {\bibinfo {volume} {983}},\ \bibinfo {pages} {A6} (\bibinfo {year} {2024})}\BibitemShut {NoStop}%
\bibitem [{\citenamefont {Ahmad}\ \emph {et~al.}(2021)\citenamefont {Ahmad}, \citenamefont {Rahimi}, \citenamefont {Tsotsas}, \citenamefont {Prat},\ and\ \citenamefont {Kharaghani}}]{ahmad_micro-scale_2021}%
  \BibitemOpen
  \bibfield  {author} {\bibinfo {author} {\bibfnamefont {F.}~\bibnamefont {Ahmad}}, \bibinfo {author} {\bibfnamefont {A.}~\bibnamefont {Rahimi}}, \bibinfo {author} {\bibfnamefont {E.}~\bibnamefont {Tsotsas}}, \bibinfo {author} {\bibfnamefont {M.}~\bibnamefont {Prat}}, \ and\ \bibinfo {author} {\bibfnamefont {A.}~\bibnamefont {Kharaghani}},\ }\href {\doibase 10.1016/j.ijheatmasstransfer.2020.120722} {\bibfield  {journal} {\bibinfo  {journal} {International Journal of Heat and Mass Transfer}\ }\textbf {\bibinfo {volume} {165}},\ \bibinfo {pages} {120722} (\bibinfo {year} {2021})}\BibitemShut {NoStop}%
\bibitem [{\citenamefont {Qin}\ \emph {et~al.}(2023)\citenamefont {Qin}, \citenamefont {Fei}, \citenamefont {Zhao}, \citenamefont {Kang}, \citenamefont {Derome},\ and\ \citenamefont {Carmeliet}}]{qin_lattice_2023}%
  \BibitemOpen
  \bibfield  {author} {\bibinfo {author} {\bibfnamefont {F.}~\bibnamefont {Qin}}, \bibinfo {author} {\bibfnamefont {L.}~\bibnamefont {Fei}}, \bibinfo {author} {\bibfnamefont {J.}~\bibnamefont {Zhao}}, \bibinfo {author} {\bibfnamefont {Q.}~\bibnamefont {Kang}}, \bibinfo {author} {\bibfnamefont {D.}~\bibnamefont {Derome}}, \ and\ \bibinfo {author} {\bibfnamefont {J.}~\bibnamefont {Carmeliet}},\ }\href {\doibase 10.1017/jfm.2023.344} {\bibfield  {journal} {\bibinfo  {journal} {Journal of Fluid Mechanics}\ }\textbf {\bibinfo {volume} {963}},\ \bibinfo {pages} {A26} (\bibinfo {year} {2023})}\BibitemShut {NoStop}%
\bibitem [{\citenamefont {Mitrovic}(2012)}]{mitrovic_josef_2012}%
  \BibitemOpen
  \bibfield  {author} {\bibinfo {author} {\bibfnamefont {J.}~\bibnamefont {Mitrovic}},\ }\href {\doibase 10.1016/j.ces.2012.03.034} {\bibfield  {journal} {\bibinfo  {journal} {Chemical Engineering Science}\ }\textbf {\bibinfo {volume} {75}},\ \bibinfo {pages} {279} (\bibinfo {year} {2012})}\BibitemShut {NoStop}%
\bibitem [{\citenamefont {Le~Dizès~Castell}\ \emph {et~al.}(2023)\citenamefont {Le~Dizès~Castell}, \citenamefont {Prat}, \citenamefont {Jabbari~Farouji},\ and\ \citenamefont {Shahidzadeh}}]{le_dizes_castell_is_2023}%
  \BibitemOpen
  \bibfield  {author} {\bibinfo {author} {\bibfnamefont {R.}~\bibnamefont {Le~Dizès~Castell}}, \bibinfo {author} {\bibfnamefont {M.}~\bibnamefont {Prat}}, \bibinfo {author} {\bibfnamefont {S.}~\bibnamefont {Jabbari~Farouji}}, \ and\ \bibinfo {author} {\bibfnamefont {N.}~\bibnamefont {Shahidzadeh}},\ }\href {\doibase 10.1021/acs.langmuir.3c00169} {\bibfield  {journal} {\bibinfo  {journal} {Langmuir}\ }\textbf {\bibinfo {volume} {39}},\ \bibinfo {pages} {5462} (\bibinfo {year} {2023})}\BibitemShut {NoStop}%
\end{thebibliography}%

 \end{document}